\documentclass[journal,draftclsnofoot,onecolumn,12pt]{IEEEtran}

\usepackage{amssymb}
\usepackage{gensymb}
\usepackage{graphicx}
\usepackage{amsmath}
\usepackage{float}
\usepackage{algorithm}
\usepackage{algpseudocode}
\usepackage{makecell}
\usepackage{cite}
\usepackage{xcolor}
\usepackage{threeparttable}
\usepackage{bm}
\usepackage{tabularx}

\usepackage{geometry}
\ifCLASSINFOpdf
\else
\fi
\begin{document}
%
\title{Rate-Splitting Multiple Access for Satellite-Terrestrial Integrated Networks: Benefits of Coordination and Cooperation}
%
%
%

\author{Longfei~Yin,~
        Bruno~Clerckx,~\IEEEmembership{Senior Member,~IEEE}
        
        \thanks{
L. Yin and B. Clerckx are with the Communications and Signal Processing Group, Department of Electrical and Electronic Engineering, Imperial College London, London, SW7 2AZ, U.K (e-mail: longfei.yin17@imperial.ac.uk; b.clerckx@imperial.ac.uk).
This work has been partially supported by the U.K. Engineering and Physical Sciences Research Council (EPSRC) under grant EP/R511547/1. }

}

\maketitle
\begin{abstract}

This work studies the joint beamforming design problem of achieving max-min rate fairness in a satellite-terrestrial integrated network (STIN) where the satellite provides wide coverage to multibeam multicast satellite users (SUs), and the terrestrial base station (BS) serves multiple cellular users (CUs) in a densely populated area. Both the satellite and BS operate in the same frequency band. Since rate-splitting multiple access (RSMA) has recently emerged as a promising strategy for non-orthogonal transmission and robust interference management in multi-antenna wireless networks, we present two RSMA-based STIN schemes, namely the \textit{coordinated scheme} relying on channel state information (CSI) sharing and the \textit{cooperative scheme} relying on CSI and data sharing. Our objective is to maximize the minimum fairness rate amongst all SUs and CUs subject to transmit power constraints at the satellite and the BS. A joint beamforming algorithm is proposed to reformulate the original problem into an approximately equivalent convex one which can be iteratively solved. Moreover, an expectation-based robust joint beamforming algorithm is proposed against the practical environment when satellite channel phase uncertainties are considered. Simulation results demonstrate the effectiveness and robustness of our proposed RSMA schemes for STIN, and exhibit significant performance gains compared with various traditional transmission strategies.

\end{abstract}


%
\IEEEpeerreviewmaketitle

\section{Introduction}
In recent years, due to the explosive growth of wireless applications and multimedia services,
satellite-terrestrial integrated network (STIN) has
gained a tremendous amount of attention in both academia and industry
as it can provide ubiquitous coverage and convey rich multimedia services, e.g., video on demand (VoD) streaming and TV broadcasting, etc. to users in both densely and sparsely populated areas\cite{kawamoto2014prospects}.
The integration of terrestrial and satellite networks is of great potential in achieving geographic coverage,
especially for remote areas where no terrestrial base station (BS) infrastructure can be employed \cite{zhang2019multicast}.
It is envisaged that the C-band ($4-8\ \mathrm{GHz}$) and S-band ($2-4\ \mathrm{GHz}$) can be shared between the terrestrial and satellite networks.
In addition, the spectrum above Ka band from $20\ \mathrm{GHz}$ to $90\ \mathrm{GHz}$ is foreseen to be the most promising candidate radio band for the next generation terrestrial cellular networks, and part of this band has already been allocated to the satellite networks \cite{maleki2015cognitive, lin2018joint}.
The concept of STIN has been proposed in the literature \cite{vassaki2013power, jia2016broadband,choi2015challenges}.
The satellite sub-network 
shares the same radio frequency band with the terrestrial sub-network through dynamic spectrum access technology
to enhance spectrum utilization, thereby achieving higher spectrum efficiency and throughput.
However, aggressive frequency reuse can induce severe interference in and between the sub-networks.
In this work, we will concentrate on RSMA-based joint beamforming schemes to efficiently mitigate the interference of STIN.

\subsection{Related Works}
A number of research efforts have investigated STIN systems.
A coexistence framework of
the satellite and terrestrial network was presented in \cite{sharma2013transmit} with the satellite link as primary and the terrestrial link as secondary.
Transmit beamforming techniques were studied to maximize the signal to interference plus noise ratio (SINR) towards secondary terrestrial users and minimize the interference towards primary satellite users.
In \cite{an2016outage}, closed-form outage probabilities of the terrestrial users
were derived while satisfying the interference constraints of satellite users.
\cite{zhu2018cooperative} studied a time division cooperative STIN, where a weighted max–min fair (MMF) beamforming design problem was formulated to jointly optimize the beamforming vectors of BSs and the satellite.
A multicast beamforming STIN system was investigated in \cite{zhang2019joint} with the aim of maximizing the sum of user minimum ratio under constraints of backhaul links and quality of service (QoS).
All aforementioned works generally assumed the satellite channels as Rician channels.
The effects of satellite antenna gain, path loss and atmospheric attenuation can be taken into account to model more practical satellite channels so as to evaluate the system performance more accurately.
In this regard, \cite{lin2018joint} investigated a joint beamforming scheme for secure communication of STIN operating in mmWave frequencies. 
Total transmit power was minimized while satisfying both the QoS constraints of a single terrestrial user and the secrecy rate (SR) requirements of multiple satellite users.
\cite{lin2018joint1} focused on the joint optimization for wireless information and power transfer (WIPT) technique in STINs.
In \cite{li2020energy}, the cache-enabled LEO satellite network was introduced, and the scheme of STIN was proposed to enable an energy-efficient radio access network (RAN) by offloading traffic from BSs through satellite’s broadcast transmission.
\cite{sharma2020secure} investigated a novel secure 3D mobile unmanned aerial vehicle (UAV) relaying for STIN 
considering three opportunistic UAV relay selection strategies.

The above works consider  conventional  space  division  multiple  access  (SDMA) based on linear precoding and assume perfect channel state information at the transmitters (CSIT). 
Each user decodes its desired stream while treating all the other interference streams as noise. 
The spatial degrees of freedom provided by multiple antennas are exploited, however the effectiveness of beamforming design relies on the accuracy of CSIT significantly.
In the real satellite communication (SatCom) environment, one practical issue is that accurate CSI is very difficult to acquire at the gateway (GW) because of the 
long-distance propagation delay and device mobility.
Thus, robust design in the presence of imperfect CSIT has been widely studied in the literature \cite{lu2019robust, guo2019robust, lin2019robust, gharanjik2015robust, wang2021resource, chu2020robust}.
\cite{lu2019robust, guo2019robust, lin2019robust} assumed the satellite channel uncertainty as additive
estimation error located in a bounded error region.
Robust beamforming was designed based on the optimization of the worst case situation.
Yet, due to the special characteristics of satellite channels, the channel magnitude does not vary significantly due to the fact that the channel propagation is dominated by the line-of-sight component.
The phase variations constitute the major source of channel uncertainty \cite{vazquez2016precoding}.
Therefore, in \cite{gharanjik2015robust, wang2021resource, chu2020robust}, robust beamforming was studied when considering constant channel amplitudes within the coherence time interval and independent time varying phase components.
Apart from the difficulties in obtaining perfect CSIT, another consideration for beamformer design in STIN is the frame-based structure of multibeam satellite standards such as DVB-S2 \cite{DVB-S2} and DVB-S2X \cite{DVB-S2x}.
Each spot beam of the satellite serves more than one user simultaneously by transmitting a single coded frame. 
Multiple users within the same beam share the same precoding vector, and leads to multibeam multicast transmission, which is a promising solution for the rapidly growing content-centric applications including video streaming, advertisements, large scale system updates and localized services, etc.

Motivated by the above practical issues in STIN, advanced multiple access techniques and interference management strategies are required.
Rate-splitting multiple access (RSMA) has recently emerged as
a promising non-orthogonal transmission for downlink multi-antenna wireless networks owning to its capability to enhance the system performance in a wide range of network loads, user deployments and CSIT qualities.
RSMA was shown analytically in \cite{clerckx2019rate} to
generalize four seemingly different strategies, namely space division multiple access (SDMA), power-domain non-orthogonal multiple access (NOMA),
orthogonal multiple access (OMA) and physical-layer multicasting.
The key behind the flexibility and robust manner of RSMA is to split each message into a common part and a private part. 
All common parts are jointly encoded into a common stream to be decoded by all users, while the private parts are individually encoded into private streams. 
The common stream is decoded at first and subtracted by successive interference cancellation (SIC). 
Each user then decodes its desired private stream and treats the remaining interference as noise. Such framework enables RSMA to partially decode interference and partially treat interference as noise. 
The benefits achieved by RSMA have been demonstrated in various multi-antenna terrestrial scenarios, such as multiuser unicast systems with perfect
CSIT \cite{clerckx2019rate, mao2018rate, ahmad2019interference,zhang2019cooperative} and imperfect CSIT \cite{joudeh2016sum, hao2015rate, joudeh2016robust, piovano2017optimal, mao2020beyond}, and multigroup multicast systems \cite{joudeh2017rate, yalcin2020rate, tervo2018multigroup}.
The superior performance of RSMA can also be seen in massive MIMO systems with residual transceiver hardware impairments \cite{papazafeiropoulos2017rate}, mmWave communications \cite{dai2017multiuser}, simultaneous wireless information and power transfer (SWIPT) networks \cite{mao2019rate} and radar-communication (RadCom) systems \cite{xu2020rate}.

More recently, the use of RSMA in multibeam SatCom or integrated satellite systems has been investigated.
\cite{caus2018exploratory} studied  RSMA  in  a two-beam  satellite  system  adopting time division multiplexing (TDM) in  each  beam. \cite{vazquez2018rate} focused  on the  sum-rate  optimization  and  low  complexity  RSMA  precoding  design by decoupling the design of common stream and private streams.
\cite{yin2020rate1, yin2020rate} proposed a RSMA-based multibeam  multicast  beamforming scheme and formulated a per-feed  power  constrained  max-min fair (MMF) problem with different CSIT qualities.
Simulation results demonstrated the effectiveness of adopting RSMA to manage inter-beam interference, taking into account various practical challenges.
In \cite{si2021rate}, RSMA was proven to be
promising for multigateway multibeam satellite systems with feeder link
interference.
\cite{lin2021supporting} considered a satellite and aerial integrated network comprising a satellite and a unmanned aerial vehicle (UAV).
The satellite employed multicast transmission, while the UAV used RSMA to improve spectral efficiency.
In \cite{lin2020secure}, a secure beamforming scheme for STIN was presented, where the satellite served one earth station (ES) with $K$ eavesdroppers (Eves).
RSMA was employed at the BS
to achieve higher spectral efficiency.
A robust beamforming scheme was proposed to maximize the secrecy energy efficiency of the ES considering Euclidean norm bounded channel uncertainty.

\subsection{Contributions}
Motivated by the prior works, we further investigate the application of RSMA into STIN consisting of a satellite sub-network and a terrestrial sub-network to manage the interference in and between both sub-networks. 
Practical  challenges are taken into account, such  as the per-feed  constraints, CSIT uncertainty,  and multibeam multicast transmission due to the existing satellite communication standards \cite{DVB-S2,DVB-S2x}.
The contributions of this article
are summarized as follows.

\begin{itemize}
\item 
{
First, we present a general framework of STIN  where  
the satellite provides wide coverage to multibeam multicast satellite users (SUs), while the terrestrial BS serves multiple cellular users (CUs) in a densely populated area.
The satellite and BS share the same radio spectrum resources, and
the GW operates as a control center to
collect and manage various kinds of information, implement centralized processing 
and control the whole system through resource allocation and interference management.
The BS is connected to the GW through optical links. 
We introduce two scenarios of RSMA-based STIN, namely the \textit{coordinated scheme} with CSI of the whole system available at the GW, and the \textit{cooperative scheme} with both CSI and transmit data information of the whole system available at the GW.
This framework differs from the prior RSMA and STIN papers, where the benefits of coordination and cooperation are not investigated.
}

\item
{
Second, based on the \textit{coordinated scheme} and \textit{cooperative scheme}, we respectively formulate optimization problems to maximize the minimum fairness rate amongst all SUs and CUs subject to the per-feed power constraints at the satellite and sum power constraint at the BS.
Perfect CSI is assumed at the GW. 
This  is  the first  work on  the joint beamforming design  for RSMA-based \textit{coordinated STIN} and \textit{cooperative STIN} with the objective of achieving max-min fairness.
Since the formulated problem is nonconvex, 
we exploit the sequential convex approximation (SCA) 
method
to reformulate the original problem into an approximately equivalent convex one.
Simulation results confirm the superiority of the \textit{cooperative scheme} over the \textit{coordinated scheme}, and 
significant performance gains of RSMA-based beamforming compared with traditional techniques.
}

\item
{
Third, since
it is very challenging to obtain accurate CSI at the GW in the practical STIN environment,
we investigate expectation-based robust beamforming design 
for both RSMA-based \textit{coordinated STIN} and \textit{cooperative STIN} 
with the satellite channel phase uncertainty taken into account.
Nonconvex MMF problems are formulated. 
A novel iterative algorithm is proposed using the SCA together with a penalty function method to tackle the optimization.
Compared with existing literatures, this is the first work to design a robust RSMA-based joint beamforming scheme in STIN considering the existence of satellite channel phase uncertainty and multibeam multicast transmission.
The effectiveness of our proposed RSMA schemes for STIN is verified through simulations.
}

\end{itemize}

The rest of this paper is organized as follows. 
The system model, channel models and two schemes of RSMA-based STIN are introduced in Section II. 
Section III investigates the design of a joint beamforming scheme to achieve max-min fairness in the STIN with the assumption of perfect CSIT.
Section IV concentrates on a robust jont beamforming scheme in the presence of satellite channel phase uncertainty.
Simulation results are given in Section V to evaluate the effectiveness of our proposed algorithms. Finally, we conclude this work in Section VI.

Notations: In the remainder of this paper, boldface uppercase, boldface lowercase and standard letters denote matrices, column vectors, and scalars respectively. 
$\mathbb{R}$ and $\mathbb{C}$ denote the real and complex domains. 
$\mathbb{E}\left ( \cdot  \right )$ is the expectation of a random variable. 
The operators $\left ( \cdot  \right )^{T}$ and $\left ( \cdot  \right )^{H}$ denote the transpose and the Hermitian transpose. $\left | \cdot  \right |$ and $\left \| \cdot  \right \|$ denote the absolute value and Euclidean norm respectively.

%
%
%
%

 


\section{System Model}
\begin{figure}
\vspace{-0.5cm}
\centering
\includegraphics[width=10.5cm]{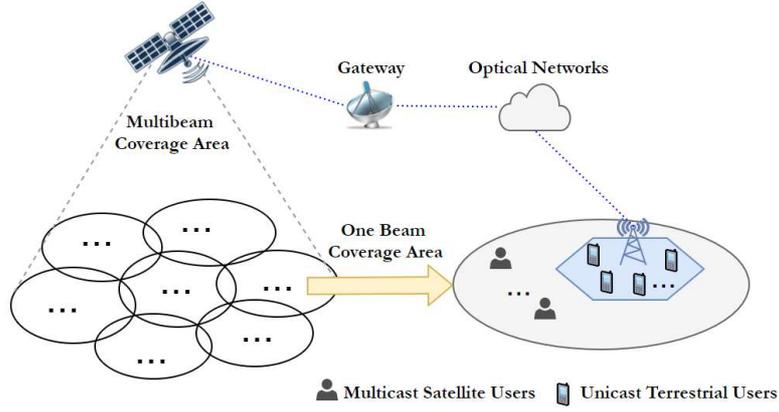}
\caption{Model of a satellite-terrestrial integrated network.}
\label{fig:stin_systemx}
\end{figure}

As illustrated in Fig. \ref{fig:stin_systemx}, we consider a STIN system
employing full frequency reuse where all SUs and CUs operate in the same frequency band.
A geostationary orbit (GEO) satellite is equipped with an array fed reflector antenna.
It provides services to the SUs that lack terrestrial access in sparsely populated or remote areas. 
By assuming a single feed per beam (SFPB) architecture, the array fed reflector antenna comprises a feed array with $N_{S}$ feeds and generates $N_{S}$ adjacent beams. 
Within the multibeam coverage area, we assume $K_{S}$ SUs, and $\rho = \frac{K_{s}}{N_{s}}$ users in each beam.
Since the SUs of one beam are served simultaneously by transmitting a single coded frame following the DVB-S2X technology, the GEO satellite implements multibeam multicast transmission.
Meanwhile, the terrestrial BS equipped with $N_{t}$-antenna uniform planar array (UPA) serves densely populated areas in the same frequency band.
$K_{t}$ unicast cellular users are assumed.
Spectrum sharing is able to improve spectrum efficiency, which also leads to interference in and between the terrestrial and satellite sub-networks.
As shown in Fig. \ref{fig:stin_systemx},
the GW acts as a control center to
collect and manage various kinds of  information, implement  centralized  processing  and  control the  whole STIN.
Optimal resource allocation and interference management on the satellite and BS can be jointly implemented at the GW to improve system performance.

\vspace{-0.3cm}
\subsection{Satellite Channel Model}
Considering the free space loss, radiation pattern and rain attenuation of satellite channels, the downlink channel vector from the satellite to SU-$k_{s}$ can be expressed as
\begin{align}
\mathbf{f}_{k_{s}} = \mathbf{b}_{k_{s}} \circ \mathbf{q}_{k_{s}} \circ  \mathrm{exp}\left \{ j \bm{\phi} _{k_{s}} \right \},
\end{align}
where 
$\circ$ is the Hadamard product.
$\mathbf{b}_{k_{s}}$
is composed of free space loss and satellite beam radiation pattern. 
The $n_{s}$-th element of $\mathbf{b}_{k_{s}}$ is modeled as
\begin{align}
b_{k_{s},n_{s}} = \frac{\sqrt{G_{R}G_{k_{s},n_{s}}}}{4\pi \frac{d_{k_{s}}}{\lambda}\sqrt{\kappa T_{sys}B_{w} }},
\label{noise_norm}
\end{align}
where $G_{R}$ is the user terminal antenna gain, $d_{k_{s}}$ is the distance between SU-$k_{s}$ and the satellite, $\lambda$ is the carrier wavelength, $\kappa$ is the Boltzmann constant, $T_{sys}$ is the receiving system noise temperature and $B_{w}$ denotes the bandwidth. 
In this regard, all channels are normalized with respect to the noise standard deviation.
The beam gain from the feed-$n_{s}$ to SU-$k_{s}$ can be approximated by
\begin{align}
G_{k_{s},n_{s}} = G_{max}\Big [\frac{J_{1}\left ( u_{k_{s},n_{s}} \right )}{2u_{k_{s},n_{s}}} + 36\frac{J_{3}\left ( u_{k_{s},n_{s}} \right )}{u_{k_{s},n_{s}}^{3}} \Big ]^{2},
\end{align}
where $u_{k_{s},n_{s}} = 2.07123\sin \left ( \theta _{k_{s},n_{s}} \right )/\sin \left ( \theta _{\mathrm{3dB}} \right )$. 
Given the $k_{s}$-th SU position,  $\theta _{k_{s},n_{s}}$ is the angle between it and the center of beam-$n_{s}$ with respect to the satellite, and $\theta _{\mathrm{3dB}}$ is the $3\ \mathrm{dB}$ loss angle compared with the beam center.
The maximum beam gain observed at each beam center is denoted by $G_{max}$.
$J_{1}$ and $J_{3}$ are respectively the first-kind Bessel functions with order $1$ and $3$.
The rain attenuation effect is characterized in $\mathbf{q}_{k_{s}} = \left [ q_{k_{s},1},q_{k_{s},2},\cdots,q_{k_{s},N_{S}} \right ]^{T}$ with elements $q_{k_{s},n_{s}} = \chi_{k_{s},n_{s}}^{1/2}$, where the dB form
$\chi_{k_{s},n_{s}}^{dB}=20\log_{10}\left ( \chi_{k_{s},n_{s}} \right )$ is commonly modeled as a log-normal random variable, i.e., $\ln \small (\chi_{k_{s},n_{s}}^{dB} \small )\sim  \mathcal{N}\left ( \mu,\sigma   \right )$. 
Moreover, $\bm{\phi}_{k_{s}} = \left [ \phi_{k_{s},1},\phi_{k_{s},2},\cdots,\phi_{k_{s},N_{S}} \right ]^{T}$ is a $N_{s}$-dimensional phase vector with each element uniformly distributed between $0$ and $2\pi$.
The satellite channel matrix between the satellite and all SUs is denoted by
$\mathbf{F} = \left [ \mathbf{f}_{1}, \cdots, \mathbf{f}_{K_{s}} \right ] \in \mathbb{C}^{N_{s} \times K_{s}}$.
Similarly, when we consider  
$n_{s} \in \left \{ 1, \cdots, N_{s} \right \}$ and $k_{t} \in \left \{ 1, \cdots, K_{t} \right \}$,
the interfering channel matrix between the satellite and all CUs 
is formulated by
$\mathbf{Z} = \left [ \mathbf{z}_{1}, \cdots, \mathbf{z}_{K_{t}} \right ] \in \mathbb{C}^{N_{s} \times K_{t}}$.

\begin{figure}
\vspace{-0.5cm}
\centering
\includegraphics[width=5.5cm]{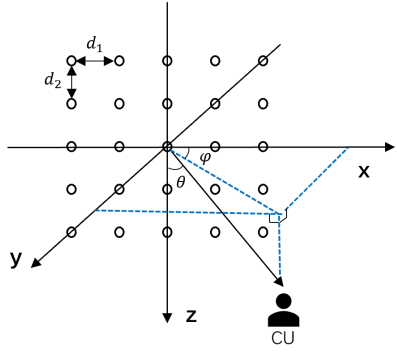}
\caption{Geometry of uniform planar
array employed at the BS.}
\label{fig:array}
\end{figure}

\subsection{Terrestrial Channel Model}
As illustrated in Fig. \ref{fig:array}, we assume UPA at the BS with dimension $N_{t} = N_{1}\times N_{2}$. 
$N_{1}$ and $N_{2}$ are respectively the number of array elements uniformly employed along the X-axis and the Z-axis.
Due to the characteristic of radio wave propagation at high frequency band, the terrestrial channels can be expressed by a model consisting of $L$ scatters.
Each scatter contributes to 
a single propagation path.
Mathematically, the terrestrial channel vector between the BS and CU $k_{t} \in  \left \{ 1, \cdots, K_{t} \right \}$ is given by
\begin{align}
    \mathbf{h}_{k_{t}} 
= \sqrt{\frac{1}{L}}\sum_{l = 1}^{L}\alpha_{k_{t},l} \mathbf{a}_{h}\left ( \theta_{k_{t},l},\varphi_{k_{t},l} \right )\otimes \mathbf{a}_{v}\left ( \theta_{k_{t},l} \right ),
\end{align}
where $\alpha_{k_{t},l}$ is the complex channel gain of the $l$-th path. 
Each $\alpha_{k_{t},l}$ is assumed to follow independent and identical distribution (i.i.d.) $\mathcal{C}\mathcal{N}\left ( 0,1 \right )$.
By denoting $\theta_{k_{t},l}$ and $\varphi_{k_{t},l}$ as the vertical and horizontal AoD angles of the $l$-th path, and $\mathbf{r}_{n_{1},n_{2}} = \left [ x_{n_{1}},0, z_{n_{2}}\right ]^{T}$ as the location vector of
the $\small (n_{1},n_{2}  \small )$-th element, the horizontal steering vector
$\mathbf{a}_{h}\left ( \theta_{k_{t},l},\varphi_{k_{t},l} \right )$ and the vertical steering vector $\mathbf{a}_{v}\small ( \theta_{k_{t},l} \small )$ can be written as
\begin{align}
    \mathbf{a}_{h}\left ( \theta_{k_{t},l},\varphi_{k_{t},l} \right ) = \big [ &e^{-j\frac{2\pi}{\lambda } \small (\frac{N_{1}-1}{2}  \small )d_{1} \sin\theta_{k_{t},l}\cos\varphi_{k_{t},l} } ,\cdots , 
    e^{+j\frac{2\pi}{\lambda } \small (\frac{N_{1}-1}{2}  \small )d_{1} \sin\theta_{k_{t},l}\cos\varphi_{k_{t},l} } \big ]^{T},
\\
     \mathbf{a}_{v}\left ( \theta_{k_{t},l}\right ) = \big [ &e^{-j\frac{2\pi}{\lambda } \small (\frac{N_{2}-1}{2}  \small )d_{2} \cos\theta_{k_{t},l} } ,\cdots , 
     e^{+j\frac{2\pi}{\lambda } \small (\frac{N_{2}-1}{2}  \small )d_{2} \cos\theta_{k_{t},l} } \big ]^{T},
\end{align}
The terrestrial channel matrix between the BS and all CUs is denoted by $\mathbf{H} = \left [ \mathbf{h}_{1}, \cdots, \mathbf{h}_{K_{t}} \right ] \in \mathbb{C}^{N_{t} \times K_{t}}$.
\subsection{Coordinated scheme and Cooperative Scheme}
In this work, we consider two levels of integration between the satellite and terrestrial BS.

\subsubsection{Coordinated Scheme}

First, we consider the basic level of integration where the CSI of both direct and interfering links of the whole network is available at the GW, while 
the transmitted data information is not shared between the satellite and BS at the GW.
We call such scheme a \textit{coordinated scheme}.
It allows the satellite and BS to coordinate power allocation and beamforming directions to suppress interference.
Different multiple access strategies can be exploited at the satellite and BS, such as RSMA, SDMA, NOMA, etc.
Here, we elaborate on the scenario where RSMA\footnote{RSMA has been shown analytically as a general multiple access strategy, which boils down to SDMA and NOMA when allocating powers to
the different types of message streams \cite{clerckx2019rate}.} is used at both the BS and satellite.
To that end, the unicast messages $W_{1},\cdots, W_{K_{t}}$ intended for CUs indexed by $\mathcal{K}_{t}=\left \{ 1, \cdots, K_{t} \right \}$ are split into common parts and private parts, 
i.e., $W_{k_{t}}\rightarrow \left \{ W_{c,k_{t}},W_{p,k_{t}} \right \}, \forall k_{t}\in \mathcal{K}_{t}$.
All common parts are combined into $W_{c}$ and encoded into a common stream $s_{c}$ to be decoded by all CUs.
The private parts are independently encoded into private streams $s_{1}, \cdots, s_{K_{t}}$.
The vector of BS streams $\mathbf{s} = \left [ s_{c}, s_{1}, \cdots, s_{K_{t}} \right ]^{T} \in \mathbb{C}^{\left ( K_{t}+1 \right ) \times 1}$ is therefore created, and we suppose it obeying $\mathbb{E}\left \{ \mathbf{s} \mathbf{s}^{H}\right \}=\mathbf{I}$.
For the satellite, multicast messages $M_{1},\cdots M_{N_{s}}$ 
are intended to the beams indexed by $\mathcal{N}_{s}=\left \{ 1, \cdots, N_{s} \right \}$.
Each message $M_{n_{s}}, \forall n_{s} \in \mathcal{N}_{s}$ is split into a common part $M_{c,n_{s}}$ and a private part $M_{p,n_{s}}$.
All common parts are combined as $M_{c}$ and encoded into $m_{c}$, while the private parts are encoded independently into $m_{1}, \cdots, m_{N_{s}}$.
The vector of satellite streams 
$\mathbf{m} = \left [ m_{c}, m_{1}, \cdots, m_{N_{s}} \right ]^{T} \in \mathbb{C}^{\left ( N_{s}+1 \right ) \times 1}$ is obtained, and we assume it satisfying $\mathbb{E}\left \{ \mathbf{m} \mathbf{m}^{H}\right \}=\mathbf{I}$.
Both $\mathbf{s}$ and $\mathbf{m}$ are linearly precoded. 
The transmitted signals at the satellite and BS are respectively
\begin{align}
    \mathbf{x}^{sat} = \mathbf{w}_{c} m_{c} + \sum_{n_{s}=1}^{N_{s}} \mathbf{w}_{n_{s}} m_{n_{s}}
\quad \mathrm{and} \quad
    \mathbf{x}^{bs} = \mathbf{p}_{c} s_{c} + \sum_{k_{t}=1}^{K_{t}} \mathbf{p}_{k_{t}} s_{k_{t}},
\end{align}
where $\mathbf{W} = \left [ \mathbf{w}_{c}, \mathbf{w}_{1}, \cdots, \mathbf{w}_{N_{s}} \right ] \in \mathbb{C}^{N_{s} \times \left ( N_{s}+1 \right )}$ and $\mathbf{P} = \left [ \mathbf{p}_{c}, \mathbf{p}_{1}, \cdots, \mathbf{p}_{K_{t}} \right ] \in \mathbb{C}^{N_{t} \times \left ( K_{t}+1 \right )}$
are defined as the beamforming matrices at the satellite and BS.
$m_{c}$ and $s_{c}$ are superimposed on top of the private signals.
Due to the lack of flexibility in sharing energy resources
among satellite feeds, per-feed transmit power constraints of the satellite are given by $\small (\mathbf{W}\mathbf{W}^{H}  \small )_{n_{s},n_{s}} \leq \frac{P_{s}}{N_{s}},\ \forall n_{s} \in \mathcal{N}_{s}$.
The sum transmit power constraint of BS is given by $\mathrm{tr}\small ( \mathbf{P} \mathbf{P}^{H} \small ) \leq P_{t}$.
Based on the channel models defined above, the received signal at each SU-$k_{s}$ writes as
\begin{equation}
    y_{k_{s}}^{sat} = \mathbf{f}_{k_{s}}^{H} \mathbf{w}_{c} m_{c}  +
    \mathbf{f}_{k_{s}}^{H} \sum_{i=1}^{N_{s}} \mathbf{w}_{i} m_{i} + n_{k_{s}}^{sat}.
\end{equation}
Each SU sees multibeam interference but no interference coming form the BS because we assume all SUs locate out of the BS serving area.
The received signal at each CU-$k_{t}$ writes as
\begin{equation}
    y_{k_{t}}^{bs} = \mathbf{h}_{k_{t}}^{H} \mathbf{p}_{c} s_{c} + \mathbf{h}_{k_{t}}^{H} \sum_{j=1}^{K_{t}} \mathbf{p}_{j} s_{j} + \mathbf{z}_{k_{t}}^{H}\mathbf{w}_{c} m_{c} + \mathbf{z}_{k_{t}}^{H} \sum_{i=1}^{N_{s}} \mathbf{w}_{i} m_{i} + n_{k_{t}}^{bs}.
\end{equation}
Each CU suffers intra-cell interference and interference from the satellite.
$\mathbf{z}_{1},\cdots, \mathbf{z}_{K{t}}$ represent satellite interfering channels.
$n_{k_{s}}^{sat}$ and $n_{k_{t}}^{bs}$ are the additive Gaussian white noises (AWGN) with zero mean and variance $\sigma _{k_{s}}^{sat2}$ and $\sigma _{k_{t}}^{bs2} $ respectively.
For both SUs and CUs,
the common stream is first decoded while treating the other interference as noise.
The SINRs of decoding the common stream at SU-$k_{s}$ and CU-$k_{t}$ are given by
\begin{align}
    \gamma_{c,k_{s}}^{sat} &= \frac{\left |\mathbf{f}_{k_{s}}^{H} \mathbf{w}_{c}  \right |^{2}}{ \sum_{i=1}^{N_{s}} \left |\mathbf{f}_{k_{s}}^{H}\mathbf{w}_{i}  \right |^{2}+ \sigma _{k_{s}}^{sat2} },
\\
    \gamma_{c,k_{t}}^{bs} &= \frac{\left |\mathbf{h}_{k_{t}}^{H} \mathbf{p}_{c}  \right |^{2}}{ \sum_{j=1}^{K_{t}} \left |\mathbf{h}_{k_{t}}^{H} \mathbf{p}_{j}  \right |^{2} + \left | \mathbf{z}_{k_{t}}^{H}\mathbf{w}_{c}  \right |^{2} + \sum_{i=1}^{N_{s}}\left | \mathbf{z}_{k_{t}}^{H} \mathbf{w}_{i} \right |^{2}  +\sigma _{k_{t}}^{bs2} }.
\end{align}
The corresponding rates are  $R_{c,k_{s}}^{sat} = \log_{2}\small ( 1+ \gamma_{c,k_{s}}^{sat} \small )$ and
$R_{c,k_{t}}^{bs} = \log_{2}\small ( 1+ \gamma_{c,k_{t}}^{bs} \small )$.
To guarantee that each SU is capable of decoding $m_{c}$, and each CU is capable of decoding $s_{c}$, we define the common rates as 
\begin{align}
    R_{c}^{sat} = \min_{k_{s}\in \mathcal{K}_{s}}\left \{ R_{c,k_{s}}^{sat} \right \}= \sum_{n_{s}=1}^{N_{s}}C_{n_{s}}^{sat}
\quad \mathrm{and} \quad
    R_{c}^{bs} = \min_{k_{t}\in \mathcal{K}_{t}}\left \{ R_{c,k_{t}}^{bs} \right \}= \sum_{k_{t}=1}^{K_{t}}C_{k_{t}}^{bs},
\end{align}
where $C_{n_{s}}^{sat}$ is the rate of the common part of the $n_{s}$-th beam's message. 
$C_{k_{t}}^{bs}$ is the rate of the common part of the $k_{t}$-th CU's message. 
After the common stream is re-encoded, precoded and subtracted from the received signal through SIC, each user then decodes its desired private stream.
We define $\mu: \mathcal{K}_{s} \rightarrow \mathcal{N}_{s}$
as mapping an SU to its corresponding beam.
The SINRs of decoding $m_{\mu\left ( k_{s} \right )}$ at SU-$k_{s}$ 
and decoding $s_{k_{t}}$ at CU-$k_{t}$ are given by
\begin{align}
    \gamma_{k_{s}}^{sat} & = \frac{\left | \mathbf{f}_{k_{s}}^{H} \mathbf{w}_{\mu\left ( k_{s} \right )} \right |^{2} }{\sum_{i=1,i \neq \mu\left ( k_{s} \right )}^{N_{s}} \left |\mathbf{f}_{k_{s}}^{H}\mathbf{w}_{i}  \right |^{2}+ \sigma _{k_{s}}^{sat2} },
\\
    \gamma_{k_{t}}^{bs} &= \frac{\left |\mathbf{h}_{k_{t}}^{H} \mathbf{p}_{k_{t}}  \right |^{2} }{ \sum_{j=1,j \neq k_{t}}^{K_{t}} \left |\mathbf{h}_{k_{t}}^{H} \mathbf{p}_{j}  \right |^{2} + \left | \mathbf{z}_{k_{t}}^{H}\mathbf{w}_{c}  \right |^{2} + \sum_{i=1}^{N_{s}}\left | \mathbf{z}_{k_{t}}^{H} \mathbf{w}_{i} \right |^{2}  +\sigma _{k_{t}}^{bs2} }.
\end{align}
The private rates are respectively 
$R_{k_{s}}^{sat}=\log_{2}\small ( 1 + \gamma _{k_{s}}^{sat} \small )$ and
$R_{k_{t}}^{bs}=\log_{2}\small ( 1 + \gamma _{k_{t}}^{bs} \small )$.
Thus, the $n_{s}$-th beam rate and the $k_{t}$-th CU rate write as
\begin{align}
    R^{sat}_{\mathrm{tot},n_{s}} =C^{sat}_{n_{s}} + \min_{i\in \mathcal{G}_{n_{s}}}R^{sat}_{i}
    \quad  \mathrm{and} \quad 
    R^{bs}_{\mathrm{tot},k_{t}} =C^{bs}_{k_{t}} + R^{bs}_{k_{t}},
\end{align}
where $\mathcal{G}_{n_{s}}$ denote the set of SUs belonging to the $n_{s}$-th beam. 
\vspace{+0.1cm}

\subsubsection{Cooperative Scheme}

Second, we consider a higher level of integration, i.e., \textit{cooperative scheme} where the satellite and BS share both CSI and data information at the GW.
In this scenario, all downlink messages
$W_{1},\cdots, W_{K_{t}}$ intended for CUs and multicast messages 
$M_{1},\cdots, M_{N_{s}}$ intended for 
SUs are transmitted at both the satellite and BS.
All propagation links (including interfering ones) are exploited to carry useful data upon appropriate beamforming. 
We still consider RSMA to manage interference in this \textit{cooperative STIN}, including inter-beam interference, intra-cell interference and interference between the satellite and terrestrial sub-networks.
Each message is split into a common part and a private part. 
All common parts are encoded together into a super common stream shared by all users in the system. 
As a result, the symbol stream to be transmitted is given by 
$\acute{\mathbf{s}}=\left [ \acute{s}_{c},\acute{m}_{1},\cdots,\acute{m}_{N_{s}},\acute{s}_{1} ,\cdots, \acute{s}_{K_{t}}\right ]^{T}\in \mathcal{C}^{N_{s}+K_{t}+1}$.
Throughout this work, we use $``\ \acute{}\ "$ to differentiate notations in the \textit{cooperative scheme} and the above \textit{coordinated scheme}.
The transmitted signals at the satellite writes as
\begin{equation}
    \acute{\mathbf{x}}^{sat} = \acute{\mathbf{w}}_{c} \acute{s}_{c} + \sum_{i=1}^{N_{s}} \acute{\mathbf{w}}_{i}^{sat} \acute{m}_{i}^{} + 
\sum_{j=1}^{K_{t}} \acute{\mathbf{w}}_{j}^{bs} \acute{s}_{j},
\end{equation}
where 
$\acute{\mathbf{W}} = \left [ \acute{\mathbf{w}}_{c}, \acute{\mathbf{w}}_{1}^{sat}, \cdots, \acute{\mathbf{w}}_{N_{s}}^{sat}, \acute{\mathbf{w}}_{1}^{bs}, \cdots, \acute{\mathbf{w}}_{K_{t}}^{bs} \right ] $ 
is the beamforming matrix, and the superscripts of  $\acute{\mathbf{w}}_{i}^{sat}$ and $\acute{\mathbf{w}}_{j}^{bs}$ are used to differentiate the precoder of satellite data and BS data.
Per-feed transmit power constraints write as $\small (\acute{\mathbf{W}}\acute{\mathbf{W}}^{H}  \small)_{n_{s},n_{s}} \leq \frac{P_{s}}{N_{s}},\ \forall n_{s} \in \mathcal{N}_{s}$.
Similarly, the transmitted signal at the BS writes as
\begin{equation}
    \acute{\mathbf{x}}^{bs} = \acute{\mathbf{p}}_{c} \acute{s}_{c} + \sum_{i=1}^{N_{s}} \acute{\mathbf{p}}_{i}^{sat} \acute{m}_{i} + 
\sum_{j=1}^{K_{t}} \acute{\mathbf{p}}_{j}^{bs} \acute{s}_{j},
\end{equation}
where 
$\acute{\mathbf{P}} = \left [ \acute{\mathbf{p}}_{c}, \acute{\mathbf{p}}_{1}^{sat}, \cdots, \acute{\mathbf{p}}_{N_{s}}^{sat}, \acute{\mathbf{p}}_{1}^{bs}, \cdots, \acute{\mathbf{p}}_{K_{t}}^{bs} \right ] $ 
is the beamforming matrix, and the sum transmit power constraint of the BS is $\mathrm{tr}\small ( \acute{\mathbf{P}}\acute{ \mathbf{P}}^{H} \small ) \leq P_{t}$.
Accordingly, the signal received at SU-$k_{s}$ becomes
\begin{equation}
   \acute{y}_{k_{s}}^{sat}=\mathbf{f}_{k_{s}}^{H} \acute{\mathbf{w}}_{c}\acute{s}_{c} +  \mathbf{f}_{k_{s}}^{H} \sum_{i=1}^{N_{s}} \acute{\mathbf{w}}_{i}^{sat} \acute{m}_{i}+
\mathbf{f}_{k_{s}}^{H} \sum _{j=1}^{K_{t}} \acute{\mathbf{w}}_{j}^{bs} \acute{s}_{j}
+ \acute{n}_{k_{s}}^{sat}.
\end{equation}
The received signal at CU-$k_{t}$ is given by
\begin{align}
    \acute{y}_{k_{t}}^{bs} &= \mathbf{h}_{k_{t}}^{H}\acute{\mathbf{p}}_{c}\acute{s}_{c} + \mathbf{h}_{k_{t}}^{H} \sum _{j = 1}^{K_{t}} \acute{\mathbf{p}}_{j}^{bs}\acute{s}_{j} + \mathbf{h}_{k_{t}}^{H} \sum _{i = 1}^{N_{s}} \acute{\mathbf{p}}_{i}^{sat}\acute{m}_{i}  + \mathbf{z}_{k_{t}}^{H}\acute{\mathbf{w}}_{c}\acute{s}_{c} + \mathbf{z}_{k_{t}}^{H} \sum _{i = 1}^{N_{s}} \acute{\mathbf{w}}_{i}^{sat}\acute{m}_{i} + \mathbf{z}_{k_{t}}^{H} \sum _{j = 1}^{K_{t}} \acute{\mathbf{w}}_{j}^{bs}\acute{s}_{j} + \acute{ n}_{k_{t}}^{bs}. \label{received signal}
\end{align}
To simplify (\ref{received signal}), aggregate channels and aggregate beamforming vectors are defined by
\begin{align}
\mathbf{g}_{k_{t}}&= \left [ \mathbf{z}_{k_{t}}^{H},\mathbf{h}_{k_{t}}^{H} \right ]^{H}\in \mathbb{C}^{\left ( N_{s}+N_{t}\right )\times 1 },\ \forall k_{t} \in \mathcal{K}_{t},
\\
\mathbf{v}_{c}&= \left [ \mathbf{w}_{c}^{*H},\mathbf{p}_{c}^{*H} \right ]^{H}\in \mathbb{C}^{\left ( N_{s}+N_{t}\right )\times 1 },
\\
\mathbf{v}_{n_{s}}^{sat} & = \left [ \acute{\mathbf{w}}_{n_{s}}^{satH},\acute{\mathbf{p}}_{n_{s}}^{satH} \right ]^{H}\in \mathbb{C}^{\left ( N_{s}+N_{t}\right )\times 1 },\ \forall n_{s} \in \mathcal{N}_{s},
\\
\mathbf{v}_{k_{t}}^{bs}&= \left [ \acute{\mathbf{w}}_{k_{t}}^{bsH},\acute{\mathbf{p}}_{k_{t}}^{bsH} \right ]^{H}\in \mathbb{C}^{\left ( N_{s}+N_{t}\right )\times 1 },\ \forall k_{t} \in \mathcal{K}_{t}.
\end{align}
The received signal at the $k_{t}$-th CU can be rewritten as
\begin{equation}
     \acute{y}_{k_{t}}^{bs}= \mathbf{g}_{k_{t}}^{H} \mathbf{v}_{c} \acute{s}_{c} + \mathbf{g}_{k_{t}}^{H} \sum _{j=1}^{K_{t}} \mathbf{v}_{j}^{bs} \acute{s}_{j}
+ \mathbf{g}_{k_{t}}^{H} \sum _{i=1}^{N_{s}} \mathbf{v}_{i}^{sat} \acute{m}_{i} + \acute{n}_{k{t}}^{bs}.
\end{equation}
From this expression, we can regard the satellite and BS working together as a super $``\mathrm{BS}"$ but subject to their respective power constraints to serve the CUs and SUs. 
Satellite interfering links are exploited to carry terrestrial data so as to improve the performance of STIN.
The aggregate beamforming vectors are collected into a matrix 
$\mathbf{V}= \small [ \mathbf{v}_{c},\mathbf{v}_{1}^{sat},\cdots, \mathbf{v}_{N_{s}}^{sat} ,\mathbf{v}_{1}^{bs},\cdots, \mathbf{v}_{K_{t}}^{bs}  \small ] \in \mathbb{C}^{\left (N_{s}+N_{t}  \right ) \times \left ( N_{s}+K_{t}+1 \right )}$.
It can also be denoted by $\mathbf{V}=\small [ \acute{\mathbf{W}}^{H} , \acute{\mathbf{P}}^{H} \small ]^{H}$.

For  both  SUs  and  CUs, the common  stream  is  firstly  decoded  and  removed  from  the  received  signal  through  SIC. 
The SINRs of decoding $\acute{s}_{c}$ at the $k_{s}$-th SU and the $k_{t}$-th CU are respectively
\begin{align}
    \acute{\gamma}_{c,k_{s}}^{sat} & = \frac{\left |\mathbf{f}_{k_{s}}^{H} \acute{\mathbf{w}}_{c}  \right |^{2}}{ \sum_{i=1}^{N_{s}} \left |\mathbf{f}_{k_{s}}^{H}\acute{\mathbf{w}}^{sat}_{i}  \right |^{2}
+ \sum_{j=1}^{K_{t}} \left |\mathbf{f}_{k_{s}}^{H}\acute{\mathbf{w}}^{bs}_{j}  \right |^{2}
+ \sigma _{k_{s}}^{sat2} },
\\
\acute{\gamma}_{c,k_{t}}^{bs} & = \frac{\left |\mathbf{g}_{k_{t}}^{H} \mathbf{v}_{c}  \right |^{2}}{ \sum_{j=1}^{K_{t}} \left |\mathbf{g}_{k_{t}}^{H} \mathbf{v}_{j}^{bs}  \right |^{2} 
+ \sum_{i=1}^{N_{s}}\left | \mathbf{g}_{k_{t}}^{H} \mathbf{v}_{i}^{sat} \right |^{2}  +\sigma _{k_{t}}^{bs2} }.
\end{align}
The corresponding rates are $\acute{R}_{c,k_{s}}^{sat} = \log_{2}\small ( 1+ \acute{\gamma}_{c,k_{s}}^{sat} \small )$ and 
$\acute{R}_{c,k_{t}}^{bs} = \log_{2}\small ( 1+ \acute{\gamma}_{c,k_{t}}^{bs} \small )$.
Since $\acute{s}_{c}$ is decoded by all users in the system, we define the common rate as 
\begin{align}
      \acute{ R}_{c} = \min_{k_{s}\in \mathcal{K}_{s}, k_{t}\in \mathcal{K}_{t}}\big \{ \acute{R}_{c,k_{s}}^{sat} ,\acute{R}_{c,k_{t}}^{bs} \big \}= \sum_{n_{s}=1}^{N_{s}}\acute{C}_{n_{s}}^{sat}+  \sum_{k_{t}=1}^{K_{t}}\acute{C}_{k_{t}}^{bs}.
\end{align}
Note that $\acute{s}_{c}$ is shared among all satellite beams and cellular users. $\acute{C}_{n_{s}}^{sat}$ and $\acute{C}_{k_{t}}^{bs}$ respectively correspond to the beam-$n_{s}$'s and CU-$k_{t}$’s portion of
common rate. 
After removing $\acute{s}_{c}$ using SIC, each user then decodes its desired private stream.
The SINRs of decoding private streams are 
\begin{align}
    \acute{\gamma}_{k_{s}}^{sat} & = \frac{\left | \mathbf{f}_{k_{s}}^{H} \acute{\mathbf{w}}^{sat}_{\mu\left ( k_{s} \right )} \right |^{2}}{ \sum_{i=1,i \neq \mu\left ( k_{s} \right )}^{N_{s}} \left |\mathbf{f}_{k_{s}}^{H}\acute{\mathbf{w}}^{sat}_{i}  \right |^{2}
+ \sum_{j=1}^{K_{t}} \left |\mathbf{f}_{k_{s}}^{H}\acute{\mathbf{w}}^{bs}_{j}  \right |^{2}
+ \sigma _{k_{s}}^{sat2} },
\\
\acute{\gamma}_{k_{t}}^{bs} & = \frac{\left |\mathbf{g}_{k_{t}}^{H} \mathbf{v}_{k_{t}}^{bs}  \right |^{2} }{\sum_{j=1,j \neq k_{t}}^{K_{t}} \left |\mathbf{g}_{k_{t}}^{H} \mathbf{v}_{j}^{bs}  \right |^{2} 
+ \sum_{i=1}^{N_{s}}\left | \mathbf{g}_{k_{t}}^{H} \mathbf{v}_{i}^{sat} \right |^{2}  +\sigma _{k_{t}}^{bs2} }.
\end{align}
$\acute{R}_{k_{s}}^{sat} = \log_{2}\small ( 1+ \acute{\gamma}_{k_{s}}^{sat} \small )$
and
$\acute{R}_{k_{t}}^{bs} = \log_{2}\small ( 1+ \acute{\gamma}_{k_{t}}^{bs} \small )$ are the private rates.
The $n_{s}$-th beam rate and the $k_{t}$-th CU rate write as
\begin{align}
    \acute{R}^{sat}_{\mathrm{tot},n_{s}} =\acute{C}^{sat}_{n_{s}} + \min_{i\in \mathcal{G}_{n_{s}}}\acute{R}^{sat}_{i}
    \quad  \mathrm{and} \quad 
    \acute{R}^{bs}_{\mathrm{tot},k_{t}} =\acute{C}^{bs}_{k_{t}} + \acute{R}^{bs}_{k_{t}}.
\end{align}

\section{Proposed Joint Beamforming Scheme}
In this section, the problem of interest is to design a joint beamforming scheme to maximize the minimum fairness rate amongst all unicast CUs and multibeam multicast SUs subject to transmit power constraints.
We respectively consider the scenarios of RSMA-based \textit{coordinated STIN} and \textit{cooperative STIN} with perfect CSI at the GW.

\subsection{Joint Beamforming Design for Coordinated STIN}
For RSMA-based \textit{coordinated STIN}, the optimization problem of achieving max-min rate fairness among all users can be formulated as
\begin{align}
     \mathcal{P}_{1}:\  \max_{\mathbf{W},\mathbf{P},\mathbf{c}^{sat},\mathbf{c}^{bs}} &\min_{n_{s}\in \mathcal{N}_{s},k_{t}\in \mathcal{K}_{t} }
    \big \{  R^{bs}_{\mathrm{tot},k_{t}} ,  R^{sat}_{\mathrm{tot},n_{s}} \big\}
    \label{p1_obj}
\\
    s.t. \qquad
    & R_{c,k_{t}}^{bs} \geq \sum_{j=1}^{K_{t}} C_{j}^{bs}, \quad \forall k_{t}\in \mathcal{K}_{t} \label{p1_constraint1}
\\
    & C_{k_{t}}^{bs} \geq 0, \quad \forall k_{t}\in \mathcal{K}_{t}
    \label{p1_constraint2}
\\
    & \mathrm{tr}\small ( \mathbf{P} \mathbf{P}^{H} \small ) \leq P_{t}
    \label{p1_constraint3}
\\
    & R_{c,k_{s}}^{sat} \geq \sum_{j=1}^{N_{s}} C_{j}^{sat}, \quad \forall k_{s}\in \mathcal{K}_{s}
    \label{p1_constraint4}
\\
    & C_{n_{s}}^{sat} \geq 0, \quad \forall n_{s}\in \mathcal{N}_{s}
    \label{p1_constraint5}
\\
    & \small ( \mathbf{W} \mathbf{W}^{H} \small )_{n_{s},n_{s}} \leq \frac{P_{s}}{N_{s}}, \quad \forall n_{s}\in \mathcal{N}_{s}
    \label{p1_constraint6}
\end{align}
where $\mathbf{c}^{sat} = \small [ C_{1}^{sat}, \cdots, C_{N_{s}}^{sat}  \small ]^{T}$, $\mathbf{c}^{bs} = \small [ C_{1}^{bs}, \cdots, C_{K_{t}}^{bs}  \small ]^{T}$ are the vectors of common rate portions.
(\ref{p1_constraint1}) guarantees that the common stream $s_{c}$ can be decoded by all CUs.
(\ref{p1_constraint3}) is the sum transmit power constraint of the BS.
Similarly, 
(\ref{p1_constraint4}) ensures the common stream $m_{c}$ to be decoded by all SUs.
(\ref{p1_constraint6}) represents per-feed power budgets of the satellite.
(\ref{p1_constraint2}) and
(\ref{p1_constraint5}) guarantee the non-negativity of all common rate portions.
Note that the formulated problem is nonconvex, we exploit an SCA-based method to convexify the nonconvex constraints and iteratively approximate the nonconvex problem to a convex one.
First, we introduce an equivalent reformulation of $ \mathcal{P}_{1}$, which is 
\begin{align}
     \mathcal{E}_{1}:\qquad  & \max_{\mathbf{W},\mathbf{P},\mathbf{c}^{sat},\mathbf{c}^{bs},q,\mathbf{r},\bm{\alpha}}  q 
    \label{e1_obj}
\\
    s.t. \qquad
    & C_{k_{t}}^{bs}+ \alpha_{k_{t}} \geq q, \quad \forall k_{t}\in \mathcal{K}_{t} 
    \label{e1_constraint1}
\\
    &  R_{k_{t}}^{bs} \geq \alpha_{k_{t}}, \quad \forall k_{t}\in \mathcal{K}_{t}
    \label{e1_constraint2}
\\
    & C_{n_{s}}^{sat}+ r_{k_{s}} \geq q, \quad \forall k_{s}\in \mathcal{G}_{n_{s}}
     \label{e1_constraint3}
\\
    & R_{k_{s}}^{sat}  \geq r_{k_{s}}, \quad \forall k_{s}\in \mathcal{K}_{s}
    \label{e1_constraint4}
\\
& (\ref{p1_constraint1})- (\ref{p1_constraint6}) \notag
\end{align}
where $q$, $\bm{\alpha }=\left [\alpha _{1}, \cdots, \alpha _{K_{t}}  \right ]^{T}$, $\mathbf{r }=\left [r_{1}, \cdots, r_{K_{s}}  \right ]^{T}$ are newly introduced auxiliary variables. 
To deal with
the nonconvexity of constraint (\ref{p1_constraint1}), (\ref{p1_constraint4}), (\ref{e1_constraint2}), (\ref{e1_constraint4}), we further introduce new auxiliary
variables $\mathbf{a }=\left [a_{1}, \cdots, a_{K_{t}}  \right ]^{T}$, $\mathbf{a }_{c}=\left [a_{c,1}, \cdots, a_{c,K_{t}}  \right ]^{T}$, $\mathbf{b }=\left [b_{1}, \cdots, b_{K_{s}}  \right ]^{T}$ and $\mathbf{b }_{c}=\left [b_{c,1}, \cdots, b_{c,K_{s}}  \right ]^{T}$.
The problem $\mathcal{E}_{1}$ is rewritten by
\begin{align}
     \mathcal{S}_{1}:\quad  & \max_{q,\mathbf{W},\mathbf{P},\mathbf{c}^{sat},\mathbf{c}^{bs},\mathbf{r},\bm{\alpha}, \mathbf{a},\mathbf{a}_{c},\mathbf{b},\mathbf{b}_{c} } q 
    \label{s1_obj}
\\
    s.t. \qquad
    &  \log \left ( 1+ a_{k_{t}} \right ) \geq \alpha_{k_{t}} \log 2, \quad \forall k_{t}\in \mathcal{K}_{t}
    \label{s1_constraint2.1}
    \\
    & \gamma_{k_{t}}^{bs} \geq a_{k_{t}}, \quad \forall k_{t}\in \mathcal{K}_{t}
    \label{s1_constraint2.2}
\\
     & \log\left ( 1+ b_{k_{s}} \right )  \geq r_{k_{s}} \log 2, \quad \forall k_{s}\in \mathcal{K}_{s}
    \label{s1_constraint4.1}
    \\
    & \gamma_{k_{s}}^{sat} \geq b_{k_{s}}, \quad \forall k_{s}\in \mathcal{K}_{s}
    \label{s1_constraint4.2}
\\
    &  \log\left ( 1+ a_{c,k_{t}} \right ) \geq \sum_{j=1}^{K_{t}} C_{j}^{bs} \log 2, \quad \forall k_{t}\in \mathcal{K}_{t}
    \label{s1_constraint5.1}
    \\
    & \gamma_{c,k_{t}}^{bs} \geq a_{c,k_{t}}, \quad \forall k_{t}\in \mathcal{K}_{t}
    \label{s1_constraint5.2}
\\
    & \log\left (  1+ b_{c,k_{s}} \right ) \geq \sum_{j=1}^{N_{s}} C_{j}^{sat} \log 2, \quad \forall k_{s}\in \mathcal{K}_{s}
    \label{s1_constraint8.1}
\\
    & \gamma_{c,k_{s}}^{sat} \geq b_{c,k_{s}}, \quad \forall k_{s}\in \mathcal{K}_{s}
    \label{s1_constraint8.2}
\\
& (\ref{p1_constraint2}), (\ref{p1_constraint3}),
(\ref{p1_constraint5}), (\ref{p1_constraint6}),
(\ref{e1_constraint1}),
(\ref{e1_constraint3})
\notag
\end{align}
where (\ref{s1_constraint2.1}) - (\ref{s1_constraint8.2}) are obtained by expanding the nonconvex rate constraints in $\mathcal{E}_{1}$.
Since the constraints of $\mathcal{S}_{1}$
hold with equality at optimality, the equivalence
between $\mathcal{P}_{1}$ and $\mathcal{S}_{1}$ can be guaranteed.
Now, we observe that the nonconvexity of $\mathcal{S}_{1}$ is due to (\ref{s1_constraint2.2}), (\ref{s1_constraint4.2}), (\ref{s1_constraint5.2}) and (\ref{s1_constraint8.2}) which contain SINR expressions.
The constraint 
(\ref{s1_constraint2.2}) can be expanded as
\begin{align}
    &\sum_{j=1,j \neq k_{t}}^{K_{t}} \left |\mathbf{h}_{k_{t}}^{H} \mathbf{p}_{j}  \right |^{2} 
    + \left | \mathbf{z}_{k_{t}}^{H}\mathbf{w}_{c}  \right |^{2} 
    + \sum_{i=1}^{N_{s}}\left | \mathbf{z}_{k_{t}}^{H} \mathbf{w}_{i} \right |^{2} 
 +\sigma _{k_{t}}^{bs2}  
    \leq \frac{\left |\mathbf{h}_{k_{t}}^{H} \mathbf{p}_{k_{t}}  \right |^{2} }{ a_{k_{t}}},
    \label{s1_constraint2.2_expand}
\end{align}
where the right-hand side (RHS) quadratic-over-linear function is convex, thus (\ref{s1_constraint2.2_expand}) is nonconvex.
Next, we need to approximate $\frac{\left |\mathbf{h}_{k_{t}}^{H} \mathbf{p}_{k_{t}}  \right |^{2} }{ a_{k_{t}}}$ with its lower bound, which is obtained by the first-order Taylor series approximation around the point $\big ( \mathbf{p}_{k_{t}}^{\left [ n \right ]} , a_{k_{t}}^{\left [ n \right ]} \big )$. Then, we have
\begin{align}
    \frac{\left |\mathbf{h}_{k_{t}}^{H} \mathbf{p}_{k_{t}}  \right |^{2} }{ a_{k_{t}}} &\geq 
\frac{2 \mathcal{R}\big ( \mathbf{p}_{k_{t}}^{\left [ n \right ] H}\mathbf{h}_{k_{t}} \mathbf{h}_{k_{t}}^{H} \mathbf{p}_{k_{t}} \big )}{a_{k_{t}}^{\left [ n \right ] }} -
\frac{\mathbf{p}_{k_{t}}^{\left [ n \right ] H}\mathbf{h}_{k_{t}} \mathbf{h}_{k_{t}}^{H} \mathbf{p}_{k_{t}}^{\left [ n \right ]}}{\big ( a_{k_{t}}^{\left [ n \right ] } \big )^{2}}a_{k_{t}} 
\triangleq 
\widehat{f_{1}}\big (\mathbf{p}_{k_{t}} , a_{k_{t}}  ;\mathbf{p}_{k_{t}}^{\left [ n \right ]} , a_{k_{t}}^{\left [ n \right ]}  \big )
\end{align}
where $n$ represents the $n$-th SCA iteration.
Replacing the linear approximation $\widehat{f_{1}}\big (\mathbf{p}_{k_{t}} , a_{k_{t}}  ;\mathbf{p}_{k_{t}}^{\left [ n \right ]} , a_{k_{t}}^{\left [ n \right ]}  \big )$ with the RHS of (\ref{s1_constraint2.2_expand}) yields
\begin{align}
    &\sum_{j=1,j \neq k_{t}}^{K_{t}} \left |\mathbf{h}_{k_{t}}^{H} \mathbf{p}_{j}  \right |^{2} + \left | \mathbf{z}_{k_{t}}^{H}\mathbf{w}_{c}  \right |^{2} + \sum_{i=1}^{N_{s}}\left | \mathbf{z}_{k_{t}}^{H} \mathbf{w}_{i} \right |^{2}  
    +\sigma _{k_{t}}^{bs2} -
    \widehat{f_{1}}\big (\mathbf{p}_{k_{t}} , a_{k_{t}}  ;\mathbf{p}_{k_{t}}^{\left [ n \right ]} , a_{k_{t}}^{\left [ n \right ]}  \big ) \leq 0.
    \label{f1}
\end{align}
Similarly, the constraint (\ref{s1_constraint4.2}) can be expanded as
\begin{align}
    \sum_{i=1,i \neq \mu\left ( k_{s} \right )}^{N_{s}} \left |\mathbf{f}_{k_{s}}^{H}\mathbf{w}_{i}  \right |^{2}+ \sigma _{k_{s}}^{sat2} \leq \frac{\left | \mathbf{f}_{k_{s}}^{H} \mathbf{w}_{\mu\left ( k_{s} \right )} \right |^{2}}{b_{k_{s}}}.
    \label{s1_constraint4.2_expand}
\end{align}
We approximate its RHS around the point $\big ( \mathbf{w}_{\mu\left ( k_{s} \right )}^{\left [ n \right ]} , b_{k_{s}}^{\left [ n \right ]} \big )$, and obtain
\begin{align}
    &\frac{\left | \mathbf{f}_{k_{s}}^{H} \mathbf{w}_{\mu\left ( k_{s} \right )} \right |^{2}}{b_{k_{s}}} \geq 
\frac{2 \mathcal{R}\big( \mathbf{w}_{\mu\left ( k_{s} \right )}^{\left [ n \right ] H}\mathbf{f}_{k_{s}} \mathbf{f}_{k_{s}}^{H} \mathbf{w}_{\mu\left ( k_{s} \right )} \big )}{b_{k_{s}}^{\left [ n \right ] }}  \notag
\\
&-
\frac{\mathbf{w}_{\mu\left ( k_{s} \right )}^{\left [ n \right ] H}\mathbf{f}_{k_{s}} \mathbf{f}_{k_{s}}^{H} \mathbf{w}_{\mu\left ( k_{s} \right )}^{\left [ n \right ]}}{\big ( b_{k_{s}}^{\left [ n \right ] } \big )^{2}}b_{k_{s}}
\triangleq 
\widehat{f_{2}}\big (\mathbf{w}_{\mu\left ( k_{s} \right )} , b_{k_{s}}  ;\mathbf{w}_{\mu\left ( k_{s} \right )}^{\left [ n \right ]} , b_{k_{s}}^{\left [ n \right ]}  \big ).
\end{align}
We replace $\widehat{f_{2}}\big (\mathbf{w}_{\mu\left ( k_{s} \right )} , b_{k_{s}}  ;\mathbf{w}_{\mu\left ( k_{s} \right )}^{\left [ n \right ]} , b_{k_{s}}^{\left [ n \right ]}  \big )$ with the RHS of (\ref{s1_constraint4.2_expand}) and yield
\begin{align}
    \sum_{i=1,i \neq \mu\left ( k_{s} \right )}^{N_{s}} \left |\mathbf{f}_{k_{s}}^{H}\mathbf{w}_{i}  \right |^{2}+ \sigma _{k_{s}}^{sat2} - 
\widehat{f_{2}}\big (\mathbf{w}_{\mu\left ( k_{s} \right )} , b_{k_{s}}  ;\mathbf{w}_{\mu\left ( k_{s} \right )}^{\left [ n \right ]} , b_{k_{s}}^{\left [ n \right ]}  \big ) \leq 0.
\label{f2}
\end{align}
Following the same logic, (\ref{s1_constraint5.2}) and (\ref{s1_constraint8.2})
are respectively approximated by
\begin{align}
   & \left |\mathbf{h}_{k_{t}}^{H} \mathbf{p}_{k_{t}}  \right |^{2} + \sum_{j=1,j \neq k_{t}}^{K_{t}} \left |\mathbf{h}_{k_{t}}^{H} \mathbf{p}_{j}  \right |^{2} + \left | \mathbf{z}_{k_{t}}^{H}\mathbf{w}_{c}  \right |^{2} + \sum_{i=1}^{N_{s}}\left | \mathbf{z}_{k_{t}}^{H} \mathbf{w}_{i} \right |^{2}  +\sigma _{k_{t}}^{bs2} -  \widehat{f_{3}}\big (\mathbf{p}_{c} , a_{c,k_{t}}  ;\mathbf{p}_{c}^{\left [ n \right ]} , a_{c,k_{t}}^{\left [ n \right ]}  \big ) \leq 0,
\label{f3}
\\
&\left | \mathbf{f}_{k_{s}}^{H} \mathbf{w}_{\mu\left ( k_{s} \right )} \right |^{2} + \sum_{i=1,i \neq \mu\left ( k_{s} \right )}^{N_{s}} \left |\mathbf{f}_{k_{s}}^{H}\mathbf{w}_{i}  \right |^{2}  + \sigma _{k_{s}}^{sat2} - \widehat{f_{4}}\big (\mathbf{w}_{c} , b_{c,k_{s}}  ;\mathbf{w}_{c}^{\left [ n \right ]} , b_{c,k_{s}}^{\left [ n \right ]}  \big ) \leq 0,
\label{f4}
\end{align}
where $\widehat{f_{3}}\big (\mathbf{p}_{c} , a_{c,k_{t}}  ;\mathbf{p}_{c}^{\left [ n \right ]} , a_{c,k_{t}}^{\left [ n \right ]}  \big ) $ and $\widehat{f_{4}}\big (\mathbf{w}_{c} , b_{c,k_{s}}  ;\mathbf{w}_{c}^{\left [ n \right ]} , b_{c,k_{s}}^{\left [ n \right ]}  \big ) $ are linear lower bound expressions 
\begin{align}
    &\widehat{f_{3}}\big (\mathbf{p}_{c} , a_{c,k_{t}}  ;\mathbf{p}_{c}^{\left [ n \right ]} , a_{c,k_{t}}^{\left [ n \right ]}  \big )  \triangleq
\frac{2 \mathcal{R}\big ( \mathbf{p}_{c}^{\left [ n \right ] H}\mathbf{h}_{k_{t}} \mathbf{h}_{k_{t}}^{H} \mathbf{p}_{c} \big )}{a_{c,k_{t}}^{\left [ n \right ] }} -
\frac{\mathbf{p}_{c}^{\left [ n \right ] H}\mathbf{h}_{k_{t}} \mathbf{h}_{k_{t}}^{H} \mathbf{p}_{c}^{\left [ n \right ]}}{\big ( a_{c,k_{t}}^{\left [ n \right ] } \big )^{2}}a_{c,k_{t}},
\\
&\widehat{f_{4}}\big (\mathbf{w}_{c} , b_{c,k_{s}}  ;\mathbf{w}_{c}^{\left [ n \right ]} , b_{c,k_{s}}^{\left [ n \right ]}  \big )  \triangleq \frac{2 \mathcal{R}\big ( \mathbf{w}_{c}^{\left [ n \right ] H}\mathbf{f}_{k_{s}} \mathbf{f}_{k_{s}}^{H} \mathbf{w}_{c} \big )}{b_{c,k_{s}}^{\left [ n \right ] }} -
\frac{\mathbf{w}_{c}^{\left [ n \right ] H}\mathbf{f}_{k_{s}} \mathbf{f}_{k_{s}}^{H} \mathbf{w}_{c}^{\left [ n \right ]}}{\big( b_{c,k_{s}}^{\left [ n \right ] } \big )^{2}}b_{c,k_{s}}.
\end{align}

Although (\ref{s1_constraint2.1}), (\ref{s1_constraint4.1}), (\ref{s1_constraint5.1}) and (\ref{s1_constraint8.1}) are convex constraints and solvable through the CVX toolbox, the $\log$ terms belong to generalized nonlinear convex program, which leads to high computational complexity.
Aiming at more efficient implementation, \cite{tervo2015optimal} approximates the $\log$ constraints to a set of second-order cone (SOC) constraints, which introduce 
a great number of slack variables
and result in an increase of per-iteration complexity.
Here, we use the property that $x \log \left ( 1 + x  \right )$ is convex as in \cite{nguyen2017distributed}, and approximate (\ref{s1_constraint2.1}), (\ref{s1_constraint4.1}), (\ref{s1_constraint5.1}), (\ref{s1_constraint8.1}) without additional slack variables. 
Since $a_{k_{t}} \geq 0$, the constraint (\ref{s1_constraint2.1}) can be rewritten as
\begin{align}
    a_{k_{t}} \log \left ( 1+ a_{k_{t}} \right ) \geq a_{k_{t}} \alpha_{k_{t}} \log 2.
    \label{s1_constraint2.1_expand}
\end{align}
Its left hand side (LHS) is convex, so
we compute the first-order Taylor series approximation of $a_{k_{t}} \log \left ( 1+ a_{k_{t}} \right )$ around the point $a_{k_{t}}^{\left [ n \right ]}$ as
\begin{align}
    & a_{k_{t}} \log \left ( 1+ a_{k_{t}} \right ) 
    \geq a_{k_{t}}^{\left [ n \right ]} \log \big ( 1+ a_{k_{t}}^{\left [ n \right ]} \big ) + \big ( a_{k_{t}} - a_{k_{t}}^{\left [ n \right ]} \big ) \notag
    \\
    & \Big [  \frac{a_{k_{t}}^{\left [ n \right ]}}{1 + a_{k_{t}}^{\left [ n \right ]}} + \log \big (1 + a_{k_{t}}^{\left [ n \right ]}  \big )   \Big ]
= a_{k_{t}} v_{k_{t}}^{\left [ n \right ]} - u_{k_{t}}^{\left [ n \right ]} ,
\end{align}
where $ v_{k_{t}}^{\left [ n \right ]}$ and $ u_{k_{t}}^{\left [ n \right ]}$ are expressions of $ a_{k_{t}}^{\left [ n \right ]}$ given by
\begin{equation}
    v_{k_{t}}^{\left [ n \right ]} = \frac{a_{k_{t}}^{\left [ n \right ]}}{a_{k_{t}}^{\left [ n \right ]}+1} + \log \big ( 1 + a_{k_{t}}^{\left [ n \right ]} \big )
    \  \mathrm{and} \ 
    u_{k_{t}}^{\left [ n \right ]} = \frac{\big (a_{k_{t}  }^{\left [ n \right ]}  \big )^{2}}{a_{k_{t}}^{\left [ n \right ]}+1} .
\end{equation}
Now, (\ref{s1_constraint2.1_expand}) can be rewritten by 
$a_{k_{t}} v_{k_{t}}^{\left [ n \right ]} - u_{k_{t}}^{\left [ n \right ]} \geq a_{k_{t}} \alpha_{k_{t}} \log 2$,
which is SOC representable as
\begin{equation}
      \Big \| \big [ a_{k_{t}}  + \alpha_{k_{t}} \log 2 - v_{k_{t}}^{\left [ n \right ]}  \quad 2\sqrt{u_{k_{t}}^{\left [ n \right ]}}\big ] \Big \|_{2}
      \leq a_{k_{t}}  - \alpha_{k_{t}} \log 2 + v_{k_{t}}^{\left [ n \right ]} .
      \label{s1_constraint2.1_transform}
\end{equation}
Similarly, the constraint (\ref{s1_constraint4.1}), (\ref{s1_constraint5.1}), (\ref{s1_constraint8.1}) can be replaced by 
\begin{align}
   & \Big \| \big [ b_{k_{s}}  + r_{k_{s}} \log 2 - \bar{v}_{k_{s}}^{\left [ n \right ]}  \quad 2\sqrt{\bar{u}_{k_{s}}^{\left [ n \right ]}}\big ] \Big \|_{2}
      \leq b_{k_{s}}  - r_{k_{s}} \log 2 + \bar{v}_{k_{s}}^{\left [ n \right ]}, \label{s1_constraint4.1_transform}
      \\
      &\Big \| \big [ a_{c,k_{t}}  + \sum_{j=1}^{K_{t}} C_{j}^{bs} \log 2 - v_{c,k_{t}}^{\left [ n \right ]}  \quad 2\sqrt{u_{c,k_{t}}^{\left [ n \right ]}}\big ] \Big \|_{2}
      \leq a_{c,k_{t}}  - \sum_{j=1}^{K_{t}} C_{j}^{bs}   \log 2 + v_{c,k_{t}}^{\left [ n \right ]},
      \label{s1_constraint5.1_transform}
      \\
      & \Big \| \big [ b_{c,k_{s}}  + \sum_{j=1}^{N_{s}} C_{j}^{sat}  \log 2 - \bar{v}_{c,k_{s}}^{\left [ n \right ]}  \quad 2\sqrt{\bar{u}_{c,k_{s}}^{\left [ n \right ]}}\big ] \Big \|_{2}
      \leq b_{c,k_{s}}  - \sum_{j=1}^{N_{s}} C_{j}^{sat}  \log 2 + \bar{v}_{c,k_{s}}^{\left [ n \right ]}. 
      \label{s1_constraint8.1_transform}
\end{align}
The expressions of $ \bar{v}_{k_{s}}^{\left [ n \right ]}$, $ \bar{u}_{k_{s}}^{\left [ n \right ]}$, $ v_{c,k_{t}}^{\left [ n \right ]}$, $ u_{c,k_{t}}^{\left [ n \right ]}$, $ \bar{v}_{c,k_{s}}^{\left [ n \right ]}$, $ \bar{u}_{c,k_{s}}^{\left [ n \right ]}$
are respectively
\begin{align}
    & \bar{v}_{k_{s}}^{\left [ n \right ]} = \frac{b_{k_{s}}^{\left [ n \right ]}}{b_{k_{s}}^{\left [ n \right ]}+1} + \log \big ( 1 + b_{k_{s}}^{\left [ n \right ]} \big )
    \ \mathrm{and} \ 
    \bar{u}_{k_{s}}^{\left [ n \right ]} = \frac{\big (b_{k_{s}  }^{\left [ n \right ]}  \big )^{2}}{b_{k_{s}}^{\left [ n \right ]}+1} , \notag \\
    & v_{c,k_{t}}^{\left [ n \right ]} = \frac{a_{c,k_{t}}^{\left [ n \right ]}}{a_{c,k_{t}}^{\left [ n \right ]}+1} + \log \big ( 1 + a_{c,k_{t}}^{\left [ n \right ]} \big )
    \  \mathrm{and} \ 
    u_{c,k_{t}}^{\left [ n \right ]} = \frac{\big (a_{c,k_{t}  }^{\left [ n \right ]}  \big )^{2}}{a_{c,k_{t}}^{\left [ n \right ]}+1} , \notag \\
    & \bar{v}_{c,k_{s}}^{\left [ n \right ]} = \frac{b_{c,k_{s}}^{\left [ n \right ]}}{b_{c,k_{s}}^{\left [ n \right ]}+1} + \log \big ( 1 + b_{c,k_{s}}^{\left [ n \right ]} \big )
    \ \mathrm{and} \ 
    \bar{u}_{c,k_{s}}^{\left [ n \right ]} = \frac{\big (b_{c,k_{s}  }^{\left [ n \right ]}  \big )^{2}}{b_{c,k_{s}}^{\left [ n \right ]}+1} .
\end{align}
Above all, the approximate convex problem at iteration-$n$ is given by
\begin{align}
     \mathcal{A}_{1}:\quad   &\max_{q,\mathbf{W},\mathbf{P},\mathbf{c}^{sat},\mathbf{c}^{bs},\mathbf{r},\bm{\alpha}, \mathbf{a},\mathbf{a}_{c},\mathbf{b},\mathbf{b}_{c} } q 
    \label{s1_obj}
\\
    s.t. \qquad 
    & 
    (\ref{f1}),(\ref{f3}),
    (\ref{s1_constraint2.1_transform}),
    (\ref{s1_constraint5.1_transform}),
    \quad \forall k_{t}\in \mathcal{K}_{t} \notag
    \\
    &
    (\ref{f2}),(\ref{f4}),
    (\ref{s1_constraint4.1_transform}),(\ref{s1_constraint8.1_transform}),
    \quad \forall k_{s}\in \mathcal{K}_{s} \notag
    \\
    & (\ref{p1_constraint2}), (\ref{p1_constraint3}),
(\ref{p1_constraint5}), (\ref{p1_constraint6}),
(\ref{e1_constraint1}),
(\ref{e1_constraint3})
\notag
\end{align}
where $\big (  \mathbf{W}^{\left [ n \right ]},\mathbf{P}^{\left [ n \right ]},\mathbf{a}^{\left [ n \right ]},\mathbf{a}_{c}^{\left [ n \right ]},\mathbf{b}^{\left [ n \right ]},\mathbf{b}_{c}^{\left [ n \right ]} \big )$ obtained from the previous iteration are treated as constants when solving $\mathcal{A}_{1}$.
Variables are updated iteratively until a stopping criterion is satisfied.
We summarize the procedure of this RSMA-based joint beamforming scheme in Algorithm \ref{algorithm1}.
$\varepsilon$ is the tolerance value.
The convergence of Algorithm \ref{algorithm1} is guaranteed since the solution of Problem $\mathcal{A}_{1}$ at iteration-$n$ is a feasible solution of the problem at iteration-$n+1$. 
As a consequence, the objective variable $q$ increases monotonically and it is bounded above by the transmit power constraints.
The solution obtained at each iteration satisfies the KKT optimality conditions of $\mathcal{A}_{1}$, which are indeed identical to those of $\mathcal{P}_{1}$ at convergence \cite{marks1978general}.

\begin{algorithm}
\caption{Proposed Joint Beamforming Scheme}\label{algorithm1}
\begin{algorithmic}
\State \textbf{Initialize}: $n\leftarrow 0,\ \mathbf{W}^{\left [ n \right ]},\mathbf{P}^{\left [ n \right ]},\mathbf{a}^{\left [ n \right ]},\mathbf{a}_{c}^{\left [ n \right ]},\mathbf{b}^{\left [ n \right ]},\mathbf{b}_{c}^{\left [ n \right ]},q^{\left [ n \right ]}$;
\Repeat
\State Solve $\mathcal{A}_{1}$ at 
$\big (  \mathbf{W}^{\left [ n \right ]},\mathbf{P}^{\left [ n \right ]},\mathbf{a}^{\left [ n \right ]},\mathbf{a}_{c}^{\left [ n \right ]},\mathbf{b}^{\left [ n \right ]},\mathbf{b}_{c}^{\left [ n \right ]} \big )$ to get

the optimal solution
$\big (  \breve{\mathbf{W}},\breve{\mathbf{P}},\breve{\mathbf{a}},\breve{\mathbf{a}}_{c},\breve{\mathbf{b}},\breve{\mathbf{b}}_{c},\breve{q} \big )$;

\State $n\leftarrow n+1$; 
\State Update $ \mathbf{W}^{\left [ n\right ]}\leftarrow \breve{\mathbf{W}},\mathbf{P}^{\left [ n \right ]}\leftarrow \breve{\mathbf{P}},\mathbf{a}^{\left [ n \right ]}\leftarrow \breve{\mathbf{a}},\mathbf{a}_{c}^{\left [ n \right ]}\leftarrow \breve{\mathbf{a}}_{c},\mathbf{b}^{\left [ n \right ]}\leftarrow \breve{\mathbf{b}},\mathbf{b}_{c}^{\left [ n \right ]} \leftarrow \breve{\mathbf{b}}_{c}, 
q^{\left [ n\right ]}\leftarrow \breve{q};$
\Until{$\left |q^{\left [ n \right ]}-q^{\left [ n -1\right ]}  \right |< \varepsilon$ ;}
\end{algorithmic}
\end{algorithm}

\vspace{-1cm}
\subsection{Joint Beamforming Design for Cooperative STIN}

When RSMA-based \textit{cooperative STIN} is considered, the optimization problem of max-min rate fairness among all users is given by
\begin{align}
\mathcal{P}_{2}:\quad
\max_{\acute{\mathbf{W}},\acute{\mathbf{P}},\acute{\mathbf{c}}} & \min_{n_{s}\in \mathcal{N}_{s},k_{t}\in \mathcal{K}_{t} }
     \big \{  \acute{R}_{k_{t}}^{bs} , \acute{R}_{k_{s}}^{sat} \big\}
    \label{p2_obj}
\\
    s.t. \qquad
    & \acute{R}_{c,k_{t}}^{bs} \geq \sum_{j=1}^{K_{t}} \acute{C}_{j}^{bs} + \sum_{j=1}^{N_{s}} \acute{C}_{j}^{sat} , \quad \forall k_{t}\in \mathcal{K}_{t}
    \label{p2_constraint1}
\\
    & \acute{C}_{k_{t}}^{bs} \geq 0 \quad \forall k_{t}\in \mathcal{K}_{t}
    \label{p2_constraint2}
\\
    & \mathrm{tr}\big ( \acute{\mathbf{P}} \acute{\mathbf{P}}^{H} \big ) \leq P_{t}
    \label{p2_constraint3}
\\
    & \acute{R}_{c,k_{s}}^{sat} \geq \sum_{j=1}^{K_{t}} \acute{C}_{j}^{bs} + \sum_{j=1}^{N_{s}} \acute{C}_{j}^{sat} , \quad \forall k_{s}\in \mathcal{K}_{s}
    \label{p2_constraint4}
\\
    & \acute{C}_{n_{s}}^{sat} \geq 0, \quad \forall n_{s}\in \mathcal{N}_{s}
    \label{p2_constraint5}
\\
    & \big ( \acute{\mathbf{W}}\acute{\mathbf{W}}^{H} \big )_{n_{s},n_{s}} \leq \frac{P_{s}}{N_{s}}, \quad \forall n_{s}\in \mathcal{N}_{s}
    \label{p2_constraint6}
\end{align}
where $\acute{\mathbf{c}}= \small [\acute{ C}_{1}^{sat}, \cdots, \acute{C}_{N_{s}}^{sat} ,\acute{ C}_{1}^{bs}, \cdots, \acute{C}_{K_{t}}^{bs} \small ]^{T}$ is the vector of all common rate portions.
(\ref{p2_constraint1}) and (\ref{p2_constraint4}) guarantee that the common stream of the whole system $\acute{s}_{c}$ can be decoded by all SUs and CUs.
(\ref{p2_constraint2}) and
(\ref{p2_constraint5}) ensure non-negativity of each element in $\acute{\mathbf{c}}$.
(\ref{p2_constraint3}) and
(\ref{p2_constraint6}) are respectively the sum transmit power constraint of the BS and per-feed transmit power constraints of the satellite.
The formulated MMF problem for \textit{cooperative STIN} is also nonconvex.
Note that the main difference between $\mathcal{P}_{1}$ and $\mathcal{P}_{2}$ lies in the transmit data information sharing in $\mathcal{P}_{2}$. 
One super common stream is transmitted at both the satellite and BS instead of transmitting individual common streams.
The expressions of rates and beamforming matrices for \textit{cooperative STIN} has been given in Section III.
We can still use the SCA-based algorithm to solve $\mathcal{P}_{2}$.
Here, we omit the detailed problem transformation and optimization framework, which follow the same procedure as that for $\mathcal{P}_{1}$.

\section{Robust Joint Beamforming Scheme}

In this section, we further investigate the robust joint beamforming scheme for RSMA-based \textit{coordinated STIN} and \textit{cooperative STIN}
considering satellite channel phase uncertainty. 
The CSIT of terrestrial channels is assumed to be perfect. 

From the satellite channel model given in Section II, 
we can observe that the amplitudes of the channel vector components are determined by some constant coefficients during the coherence time interval, including the free space loss, satellite antenna gain and rain attenuation \cite{wang2021resource}.
However, the satellite channel phases vary rapidly due to a series of time-varying factors, such as the use of different local oscillators (LO) on-board, the rain, cloud and gaseous absorption, and the use of low-noise block (LNB) at receive sides \cite{ wang2021resource, vazquez2016precoding}.
Therefore, within a coherence time interval, 
the phase of the channel vector from the satellite to SU-$k_{s}$ at time instant $t_{1}$ can be modeled as
\begin{align}
\bm{\phi}_{k_{s}}\left ( t_{1} \right ) = \bm{\phi}_{k_{s}}\left ( t_{0}  \right )+ \mathbf{e}_{k_{s}},
\end{align}
where $\bm{\phi}_{k_{s}}\left ( t_{0}  \right )$ represents the phase vector which is estimated at the previous time instant $t_{0}$ and fed back to the GW.
$\mathbf{e}_{k_{s}} = \left [ e_{k_{s},1}, e_{k_{s},2},\cdots, e_{k_{s},N_{s}} \right ]^{T}$ is the phase uncertainty following the distribution $\mathbf{e}_{k_{s}} \sim \mathcal{N}\left ( \mathbf{0},\delta^{2} \mathbf{I} \right )$, with i.i.d Gaussian random entries.
For ease of notation, we can generally indicate $\bm{\phi}_{k_{s}}\left ( t_{1} \right )$ and $\bm{\phi}_{k_{s}}\left ( t_{0} \right )$ by $\bm{\phi}_{k_{s}}$ and $\widehat{\bm{\phi}}_{k_{s}}$ respectively.
Since we assume identical channel amplitudes within the coherence time interval, the channel vector from the satellite to SU-$k_{s}$ can be written as
\begin{align}
\mathbf{f}_{k_{s}} = \widehat{\mathbf{f}}_{k_{s}} \circ \mathbf{x}_{k_{s}} = \mathrm{diag}\big (  \widehat{\mathbf{f}}_{k_{s}} \big )\mathbf{x}_{k_{s}},
\end{align}
where $\mathbf{x}_{k_{s}} = \mathrm{exp}\left \{ j \mathbf{e}_{k_{s}}\right \}$ is a random vector.
We further assume that the channel estimate $\widehat{\mathbf{f}}_{k_{s}}$ and the correlation matrix of $\mathbf{x}_{k_{s}}$ denoted by $\mathbf{X}_{k_{s}} =  \mathbb{E}\left \{  \mathbf{x}_{k_{s}} \mathbf{x}_{k_{s}} ^{H}\right \}$ are known at the GW \cite{ wang2021resource, gharanjik2015precoding}.
For the interfering channels, by defining $\mathbf{y}_{k_{t}} = \mathrm{exp}\left \{ j {\mathbf{e}}'_{k_{t}}\right \}$ and ${\mathbf{e}}'_{k_{t}} = \left [ {e}'_{k_{t},1}, {e}'_{k_{t},2},\cdots, {e}'_{k_{t},N_{s}} \right ]^{T}
$ 
following ${\mathbf{e}}'_{k_{t}} \sim \mathcal{N}\left ( \mathbf{0},\delta^{2} \mathbf{I} \right )$, 
the channel vector from the satellite to CU-$k_{t}$ write as
\begin{align}
\mathbf{z}_{k_{t}} = \widehat{\mathbf{z}}_{k_{t}} \circ \mathbf{y}_{k_{t}} = \mathrm{diag}\big (  \widehat{\mathbf{z}}_{k_{t}} \big )\mathbf{y}_{k_{t}},
\end{align}
where the channel estimate $\widehat{\mathbf{z}}_{k_{t}}$ and the correlation matrix $\mathbf{Y}_{k_{t}} =  \mathbb{E}\left \{  \mathbf{y}_{k_{t}} \mathbf{y}_{k_{t}} ^{H}\right \}$ are available at the GW.
Hence, we concentrate on the expectation-based robust beamforming design.
The MMF optimization problem for RSMA-based \textit{coordinated STIN} considering satellite phase uncertainty remains the same as $\mathcal{P}_{1}$ in Section III,
By introducing auxiliary variables $q$, $\bm{\alpha }=\left [\alpha _{1}, \cdots, \alpha _{K_{t}}  \right ]^{T}$, $\mathbf{r }=\left [r_{1}, \cdots, r_{K_{s}}  \right ]^{T}$, $W = \left \{ \mathbf{W}_{c}, \mathbf{W}_{1}, \cdots, \mathbf{W}_{N_{s}} \right \}$ and $P = \left \{ \mathbf{P}_{c}, \mathbf{P}_{1}, \cdots, \mathbf{P}_{K_{t}} \right \}$, the original $\mathcal{P}_{1}$ can be equivalently transformed into a semi-definite programming (SDP) problem with rank-one constraints
\begin{align}
     \mathcal{D}_{1}:\quad  & \max_{W,P,\mathbf{c}^{sat},\mathbf{c}^{bs},q,\mathbf{r},\bm{\alpha}}  q 
    \label{e1r_obj}
\\
    s.t. \qquad
&\mathrm{tr}\left ( \mathbf{P}_{c} \right ) + \sum_{k_{t}=1}^{K_{t}}\mathrm{tr}\left ( \mathbf{P}_{k_{t}} \right ) \leq P_{t}
\label{d1_constraint1}
\\
& \big [ \mathbf{W}_{c} + \sum_{i=1}^{N_{s}}\mathbf{W}_{i}  \big]_{n_{s},n_{s}}  \leq \frac{P_{s}}{N_{s}},  \quad n_{s} \in \mathcal{N}_{s}
\label{d1_constraint2}
\\
&\mathbf{W}_{c} \succeq 0,\ \mathbf{W}_{n_{s}} \succeq 0 , \quad \forall n_{s} \in \mathcal{N}_{s}
\label{d1_constraint3}
\\
&\mathbf{P}_{c} \succeq 0,\ \mathbf{P}_{k_{t}} \succeq 0 , \quad \forall k_{t} \in \mathcal{K}_{t}
\label{d1_constraint4}
\\
& \mathrm{rank}\left ( \mathbf{W}_{c} \right ) =1,\ \mathrm{rank}\left ( \mathbf{W}_{n_{s}} \right ) =1 , \  \forall n_{s} \in \mathcal{N}_{s}
\label{d1_constraint5}
\\
& \mathrm{rank}\left ( \mathbf{P}_{c} \right ) =1,\ \mathrm{rank}\left ( \mathbf{P}_{k_{t}} \right ) =1 , \quad \forall k_{t} \in \mathcal{K}_{t}
\label{d1_constraint6}
\\
& (\ref{p1_constraint1}),
(\ref{p1_constraint2}),
(\ref{p1_constraint4}),
(\ref{p1_constraint5}) ,
(\ref{e1_constraint1})-
(\ref{e1_constraint4}) 
\notag
\end{align}
where $\mathbf{W}_{c} = \mathbf{w}_{c}\mathbf{w}_{c}^{H}$, $\left \{ \mathbf{W}_{n_{s}} = \mathbf{w}_{n_{s}}\mathbf{w}_{n_{s}}^{H} \right \}_{n_{s}=1}^{N_{s}}$, $\mathbf{P}_{c} = \mathbf{p}_{c}\mathbf{p}_{c}^{H}$, $\left \{ \mathbf{P}_{k_{t}} = \mathbf{p}_{k_{t}}\mathbf{p}_{k_{t}}^{H} \right \}_{k_{t}=1}^{K_{t}}$.
(\ref{d1_constraint1}) and (\ref{d1_constraint1}) are the equivalent transmit power constraints.
All the rate expressions in this section are redefined by the general form
$R \triangleq \mathbb{E} \left \{  \log_{2}\left ( 1+ \mathrm{SINR} \right ) \right \}$, which is averaged with respect to the satellite channel phase uncertainty to guarantee the robustness. 
It has been verified in \cite{chu2020robust,shao2017simple} that for nonnegative random variables $A$ and $B$, we can use the following approximation
\begin{align}
    \mathbb{E} \Big \{  \log_{2}\Big ( 1+ \frac{A}{B} \Big ) \Big \} \approx \log_{2} \Big ( 1+ \frac{\mathbb{E} \left ( A \right )}{\mathbb{E} \left ( B \right)} \Big) .
\end{align}
By taking (\ref{e1_constraint4}) as an example, $R_{k_{s}}^{sat}$ can be approximated by
\begin{align}
    R_{k_{s}}^{sat} &= \mathbb{E} \left \{ \log_{2}\left ( 1+ \gamma_{k_{s}}^{sat}  \right ) \right \} \notag
    \\
    &\approx \log_{2}
 \left (  
\frac{ \mathbb{E} \left \{ \mathrm{tr} \left ( \mathbf{F}_{k_{s}}\mathbf{W}_{\mu\left ( k_{s} \right )} \right )  \right \}    + \sum_{i=1,i \neq \mu\left ( k_{s} \right )}^{N_{s}}  \mathbb{E}\left \{ \mathrm{tr}\left ( \mathbf{F}_{k_{s}} \mathbf{W}_{i} \right )  \right \}  + \sigma _{k_{s}}^{sat2}      }
{   \sum_{i=1,i \neq \mu\left ( k_{s} \right )}^{N_{s}}  \mathbb{E}\left \{ \mathrm{tr}\left ( \mathbf{F}_{k_{s}} \mathbf{W}_{i} \right )  \right \}  + \sigma _{k_{s}}^{sat2}     } 
\right ) \notag
\\
&=
\log_{2}
 \left (  
\frac{ \mathrm{tr} \left ( \overline{\mathbf{F}}_{k_{s}} \mathbf{W}_{\mu\left ( k_{s} \right )} \right )     + \sum_{i=1,i \neq \mu\left ( k_{s} \right )}^{N_{s}}   \mathrm{tr}\left ( \overline{\mathbf{F}}_{k_{s}} \mathbf{W}_{i} \right )  + \sigma _{k_{s}}^{sat2}      }
{   \sum_{i=1,i \neq \mu\left ( k_{s} \right )}^{N_{s}}   \mathrm{tr}\left ( \overline{\mathbf{F}}_{k_{s}} \mathbf{W}_{i} \right )    + \sigma _{k_{s}}^{sat2}     } 
\right ).\label{eq91}
\end{align}

Specifically, $\mathbf{F}_{k_{s}} = \mathrm{diag}\big ( \widehat{\mathbf{f}}_{k_{s}}  \big ) \mathbf{x}_{k_{s}}\mathbf{x}^{H}_{k_{s}}
\mathrm{diag}\big ( \widehat{\mathbf{f}}^{H}_{k_{s}}  \big )$, and $\overline{  \mathbf{F}}_{k_{s}} = \mathbb{E}\left \{ \mathbf{F}_{k_{s}}  \right \} = \mathrm{diag}\big ( \widehat{\mathbf{f}}_{k_{s}}  \big ) \mathbf{X}_{k_{s}}
\mathrm{diag}\big ( \widehat{\mathbf{f}}^{H}_{k_{s}}  \big )$ is defined as the channel correlation matrix which can capture the expectation over the distribution of phase uncertainty.
Based on the approximated rate expressions, $\mathcal{D}_{1}$
can be rewritten as $\mathcal{F}_{1}$.
\begin{align}
     \mathcal{F}_{1}:\quad  & \max_{W,P,\mathbf{c}^{sat},\mathbf{c}^{bs},q,\mathbf{r},\bm{\alpha}, \eta, \xi
     } q 
     \label{f1_objective}
\\
s.t. \qquad
& \eta^{bs}_{k_{t}} - \xi^{bs} _{k_{t}} \geq \alpha_{k_{t}} \log 2, \  \forall k_{t}\in \mathcal{K}_{t}
\label{f1_constraint1}
\\
& e^{\eta^{bs}_{k_{t}}} \leq \mathrm{tr}\left (\mathbf{H}_{k_{t}}\mathbf{P}_{k_{t}} \right ) + \sum_{j=1,j \neq k_{t}}^{K_{t}}\mathrm{tr}\left (\mathbf{H}_{k_{t}}\mathbf{P}_{j} \right )
+ \mathrm{tr}\left (\overline{\mathbf{Z}}_{k_{t}}\mathbf{W}_{c} \right ) + \sum_{i=1}^{N_{s}}\mathrm{tr}\left (\overline{\mathbf{Z}}_{k_{t}}\mathbf{W}_{i} \right ) + \sigma _{k_{t}}^{bs2}  , \  \forall k_{t}\in \mathcal{K}_{t}
\label{f1_constraint2}
\\
& e^{ \xi^{bs} _{k_{t}}} \geq \sum_{j=1,j \neq k_{t}}^{K_{t}}\mathrm{tr}\left (\mathbf{H}_{k_{t}}\mathbf{P}_{j} \right ) + \mathrm{tr}\left (\overline{\mathbf{Z}}_{k_{t}}\mathbf{W}_{c} \right ) + \sum_{i=1}^{N_{s}}\mathrm{tr}\left (\overline{\mathbf{Z}}_{k_{t}}\mathbf{W}_{i} \right ) + \sigma _{k_{t}}^{bs2}, \  \forall k_{t}\in \mathcal{K}_{t}
\label{f1_constraint3}
\\
& \eta^{sat} _{k_{s}} - \xi^{sat} _{k_{s}}\geq  r_{k_{s}} \log 2, \ \forall k_{s}\in \mathcal{K}_{s}
\label{f1_constraint4}
\\
& e^{ \eta^{sat} _{k_{s}}} \leq \mathrm{tr} \left ( \overline{\mathbf{F}}_{k_{s}} \mathbf{W}_{\mu\left ( k_{s} \right )} \right )     + \sum_{i=1,i \neq \mu\left ( k_{s} \right )}^{N_{s}}   \mathrm{tr}\left ( \overline{\mathbf{F}}_{k_{s}} \mathbf{W}_{i} \right )  + \sigma _{k_{s}}^{sat2}   , \ \forall k_{s}\in \mathcal{K}_{s}
\label{f1_constraint5}
\\
& e^{ \xi^{sat} _{k_{s}}} \geq \sum_{i=1,i \neq \mu\left ( k_{s} \right )}^{N_{s}}   \mathrm{tr}\left ( \overline{\mathbf{F}}_{k_{s}} \mathbf{W}_{i} \right )    + \sigma _{k_{s}}^{sat2}  , \ \forall k_{s}\in \mathcal{K}_{s}
\label{f1_constraint6}
\\
& \eta^{bs}_{c,k_{t}} - \xi^{bs} _{c,k_{t}} \geq \sum_{j=1}^{K_{t}} C_{j}^{bs} \log 2, \ \forall k_{t}\in \mathcal{K}_{t}
\label{f1_constraint7}
\\
& e^{\eta^{bs}_{c,k_{t}}} \leq \mathrm{tr}\left (\mathbf{H}_{k_{t}}\mathbf{P}_{c} \right ) + \sum_{j=1}^{K_{t}}\mathrm{tr}\left (\mathbf{H}_{k_{t}}\mathbf{P}_{j} \right )
+ \mathrm{tr}\left (\overline{\mathbf{Z}}_{k_{t}}\mathbf{W}_{c} \right ) 
+ \sum_{i=1}^{N_{s}}\mathrm{tr}\left (\overline{\mathbf{Z}}_{k_{t}}\mathbf{W}_{i} \right ) + \sigma _{k_{t}}^{bs2}  , \  \forall k_{t}\in \mathcal{K}_{t}
\label{f1_constraint8}
\\
& e^{\xi^{bs}_{c,k_{t}}} \geq \sum_{j=1}^{K_{t}}\mathrm{tr}\left (\mathbf{H}_{k_{t}}\mathbf{P}_{j} \right ) + \mathrm{tr}\left(\overline{\mathbf{Z}}_{k_{t}}\mathbf{W}_{c} \right )  + \sum_{i=1}^{N_{s}}\mathrm{tr}\left (\overline{\mathbf{Z}}_{k_{t}}\mathbf{W}_{i} \right ) + \sigma _{k_{t}}^{bs2}  , \  \forall k_{t}\in \mathcal{K}_{t}
\label{f1_constraint9}
\\
&  \eta^{sat}_{c,k_{s}} - \xi^{sat} _{c,k_{s}} \geq \sum_{j=1}^{N_{s}} C_{j}^{sat} \log 2, \ \forall k_{s}\in \mathcal{K}_{s}
\label{f1_constraint10}
\\
& e^{\eta^{sat}_{c,k_{s}}} \leq \mathrm{tr} \left ( \overline{\mathbf{F}}_{k_{s}} \mathbf{W}_{c} \right )     + \sum_{i=1}^{N_{s}}   \mathrm{tr}\left ( \overline{\mathbf{F}}_{k_{s}} \mathbf{W}_{i} \right )  + \sigma _{k_{s}}^{sat2}   , \ \forall k_{s}\in \mathcal{K}_{s}
\label{f1_constraint11}
\\
& e^{\xi^{sat} _{c,k_{s}}} \geq \sum_{i=1}^{N_{s}}   \mathrm{tr}\left ( \overline{\mathbf{F}}_{k_{s}} \mathbf{W}_{i} \right )    + \sigma _{k_{s}}^{sat2}  , \ \forall k_{s}\in \mathcal{K}_{s}
\label{f1_constraint12}
\\
& (\ref{p1_constraint2}),
(\ref{p1_constraint5}) ,
(\ref{e1_constraint1}),
(\ref{e1_constraint3}) ,
(\ref{d1_constraint1})-
(\ref{d1_constraint6})
\notag
\end{align}
where $\eta$ and $\xi$ are the sets of introduced slack variables.
The constraints (\ref{f1_constraint1})-(\ref{f1_constraint3}), (\ref{f1_constraint4})-(\ref{f1_constraint6}), (\ref{f1_constraint7})-(\ref{f1_constraint9}), and (\ref{f1_constraint10})-(\ref{f1_constraint12}) are respectively the expansions of the rate constraints  (\ref{e1_constraint2}), (\ref{e1_constraint4}), (\ref{p1_constraint1}) and (\ref{p1_constraint4}).
Note that (\ref{f1_constraint3}), (\ref{f1_constraint6}), (\ref{f1_constraint9}) and  (\ref{f1_constraint12}) are nonconvex  with convex LHSs which can be approximated by the first-order Taylor approximation.
Hence, we obtain these approximated linear constraints
\begin{align}
& \sum_{j=1,j \neq k_{t}}^{K_{t}}\mathrm{tr}\left (\mathbf{H}_{k_{t}}\mathbf{P}_{j} \right ) + \mathrm{tr}\left (\overline{\mathbf{Z}}_{k_{t}}\mathbf{W}_{c} \right ) + \sum_{i=1}^{N_{s}}\mathrm{tr}\left (\overline{\mathbf{Z}}_{k_{t}}\mathbf{W}_{i} \right )
+ \sigma _{k_{t}}^{bs2} \leq
e^{ \xi^{bs\left [ n \right ]} _{k_{t}}}\big( \xi^{bs} _{k_{t}} - \xi^{bs\left [ n \right ]} _{k_{t}} +1 \big ) ,
\label{g1_constraint1}
\\
& \sum_{i=1,i \neq \mu\left ( k_{s} \right )}^{N_{s}}   \mathrm{tr}\left ( \overline{\mathbf{F}}_{k_{s}} \mathbf{W}_{i} \right )    + \sigma _{k_{s}}^{sat2} 
\leq e^{ \xi^{sat\left [ n \right ]} _{k_{s}}}\big ( \xi^{sat} _{k_{s}} - \xi^{sat\left [ n \right ]} _{k_{s}} +1 \big ),
\label{g1_constraint2}
\\
& \sum_{j=1}^{K_{t}}\mathrm{tr}\left (\mathbf{H}_{k_{t}}\mathbf{P}_{j} \right ) + \mathrm{tr}\left(\overline{\mathbf{Z}}_{k_{t}}\mathbf{W}_{c} \right )  + \sum_{i=1}^{N_{s}}\mathrm{tr}\left (\overline{\mathbf{Z}}_{k_{t}}\mathbf{W}_{i} \right ) + \sigma _{k_{t}}^{bs2} \leq e^{\xi^{bs\left [ n \right ]}_{c,k_{t}}} \big( \xi^{bs} _{c,k_{t}} - \xi^{bs\left [ n \right ]} _{c,k_{t}} +1 \big ) ,
\label{g1_constraint3}
\\
& \sum_{i=1}^{N_{s}}   \mathrm{tr}\left ( \overline{\mathbf{F}}_{k_{s}} \mathbf{W}_{i} \right )    + \sigma _{k_{s}}^{sat2} 
\leq
e^{\xi^{sat\left [ n \right ]} _{c,k_{s}}} \big ( \xi^{sat} _{c,k_{s}} - \xi^{sat\left [ n \right ]} _{c,k_{s}} +1\big ).
\label{g1_constraint4}
\end{align}
where $n$ represents the $n$-th SCA iteration.
The constraints (\ref{f1_constraint3}), (\ref{f1_constraint6}), (\ref{f1_constraint9}) and  (\ref{f1_constraint12}) belong to generalized nonlinear convex program with high computational complexity.
Following the same method introduced in Section III, they can be represented in linear and SOC forms given by
\begin{align}
& t^{bs}_{k_{t}} \leq \mathrm{tr}\left (\mathbf{H}_{k_{t}}\mathbf{P}_{k_{t}} \right ) + \sum_{j=1,j \neq k_{t}}^{K_{t}}\mathrm{tr}\left (\mathbf{H}_{k_{t}}\mathbf{P}_{j} \right )
 + \mathrm{tr}\left (\overline{\mathbf{Z}}_{k_{t}}\mathbf{W}_{c} \right ) + \sum_{i=1}^{N_{s}}\mathrm{tr}\left (\overline{\mathbf{Z}}_{k_{t}}\mathbf{W}_{i} \right ) + \sigma _{k_{t}}^{bs2},
 \label{g1_constraint5}
 \\
 & \Big \|  t^{bs}_{k_{t}} + \eta^{bs}_{k_{t}} - \big (\log\small ( t^{bs\left [ n \right ]}_{k_{t}} \small)+1  \big ) \quad 2\sqrt{t^{bs\left [ n \right ]}_{k_{t}}} \Big \|^{2} 
\leq t^{bs}_{k_{t}} - \eta^{bs}_{k_{t}} + \big (\log\small ( t^{bs\left [ n \right ]}_{k_{t}} \small)+1  \big ) ,
\label{g1_constraint6}
\\
& t^{sat}_{k_{s}} \leq \mathrm{tr} \left ( \overline{\mathbf{F}}_{k_{s}} \mathbf{W}_{\mu\left ( k_{s} \right )} \right )     + \sum_{i=1,i \neq \mu\left ( k_{s} \right )}^{N_{s}}   \mathrm{tr}\left ( \overline{\mathbf{F}}_{k_{s}} \mathbf{W}_{i} \right )  + \sigma _{k_{s}}^{sat2} ,
\label{g1_constraint7}
\\
& \Big \|  t^{sat}_{k_{s}} + \eta^{sat}_{k_{s}} - \big (\log\small ( t^{sat\left [ n \right ]}_{k_{s}} \small)+1  \big ) \quad 2\sqrt{t^{sat\left [ n \right ]}_{k_{s}}} \Big \|^{2} 
\leq t^{sat}_{k_{s}} - \eta^{sat}_{k_{s}} + \big (\log\small ( t^{sat\left [ n \right ]}_{k_{s}} \small)+1  \big ),
\label{g1_constraint8}
\\
& t^{bs}_{c,k_{t}} \leq \mathrm{tr}\left (\mathbf{H}_{k_{t}}\mathbf{P}_{c} \right ) + \sum_{j=1}^{K_{t}}\mathrm{tr}\left (\mathbf{H}_{k_{t}}\mathbf{P}_{j} \right )
+ \mathrm{tr}\left (\overline{\mathbf{Z}}_{k_{t}}\mathbf{W}_{c} \right ) 
+ \sum_{i=1}^{N_{s}}\mathrm{tr}\left (\overline{\mathbf{Z}}_{k_{t}}\mathbf{W}_{i} \right ) + \sigma _{k_{t}}^{bs2},
\label{g1_constraint9}
\\
& \Big \|  t^{bs}_{c,k_{t}} + \eta^{bs}_{c,k_{t}} - \big (\log\small ( t^{bs \left [ n \right ]}_{c,k_{t}} \small)+1  \big ) \quad 2\sqrt{t^{bs \left [ n \right ]}_{c,k_{t}}} \Big \|^{2} 
\leq t^{bs}_{c,k_{t}} - \eta^{bs}_{c,k_{t}} + \big (\log\small ( t^{bs \left [ n \right ]}_{c,k_{t}} \small)+1  \big ),
\label{g1_constraint10}
\\
& t^{sat}_{c,k_{s}} \leq \mathrm{tr} \left ( \overline{\mathbf{F}}_{k_{s}} \mathbf{W}_{c} \right )     + \sum_{i=1}^{N_{s}}   \mathrm{tr}\left ( \overline{\mathbf{F}}_{k_{s}} \mathbf{W}_{i} \right )  + \sigma _{k_{s}}^{sat2} ,
\label{g1_constraint11}
\\
& \Big \|  t^{sat}_{c,k_{s}} + \eta^{sat}_{c,k_{s}} - \big (\log\small ( t^{sat \left [ n \right ]}_{c,k_{s}} \small)+1  \big ) \quad 2\sqrt{t^{sat \left [ n \right ]}_{c,k_{s}}} \Big \|^{2}
\leq t^{sat}_{c,k_{s}} - \eta^{sat}_{c,k_{s}} + \big (\log\small ( t^{sat \left [ n \right ]}_{c,k_{s}} \small)+1  \big ).
\label{g1_constraint12}
\end{align}

Since rank-one implies only one nonzero eigenvalue, the nonconvex constraints 
(\ref{d1_constraint5}) and (\ref{d1_constraint6})
can be rewritten by 
\begin{align}
& \mathrm{tr}\left ( \mathbf{W}_{c} \right ) - \lambda _{\max}\left ( \mathbf{W}_{c} \right ) = 0, \ \mathrm{tr}\left ( \mathbf{W}_{n_{s}} \right ) - \lambda _{\max}\left ( \mathbf{W}_{n_{s}} \right ) = 0,\ \forall n_{s}\in \mathcal{N}_{s},
\\
& \mathrm{tr}\left ( \mathbf{P}_{c} \right ) - \lambda _{\max}\left ( \mathbf{P}_{c} \right ) = 0, \ \mathrm{tr}\left ( \mathbf{P}_{k_{t}} \right ) - \lambda _{\max}\left ( \mathbf{P}_{k_{t}} \right ) = 0,\ \forall k_{t}\in \mathcal{K}_{t},
\end{align}
where $\lambda _{\max}\left ( \mathbf{X} \right )$ denotes the maximum eigenvalue of $\mathbf{X} \succeq 0$.
Then, we build a penalty function to insert  these constraints into the objective function (\ref{f1_objective})
and obtain 
\begin{align}
    \max_{W,P,\mathbf{c}^{sat},\mathbf{c}^{bs},q,\mathbf{r},\mathbf{\alpha}, \eta, \xi
     }  q & - \beta \Big (\left [\mathrm{tr}\left ( \mathbf{W}_{c} \right ) - \lambda _{\max}\left ( \mathbf{W}_{c} \right )  \right ]  + \sum_{n_{s}=1}^{N_{s}}\left [\mathrm{tr}\left ( \mathbf{W}_{n_{s}} \right ) - \lambda _{\max}\left ( \mathbf{W}_{n_{s}} \right )  \right ] \notag
     \\
     & + \left [\mathrm{tr}\left ( \mathbf{P}_{c} \right ) - \lambda _{\max}\left ( \mathbf{P}_{c} \right )  \right ]  + \sum_{k_{t}=1}^{K_{t}}\left [\mathrm{tr}\left ( \mathbf{P}_{k_{t}} \right ) - \lambda _{\max}\left ( \mathbf{P}_{k_{t}} \right )  \right ]   \Big ).
     \label{new_objective}
\end{align}
$\beta$ is a proper penalty factor to guarantee the penalty function as small as possible.
(\ref{new_objective}) is nonconcave due to the existence of the penalty function. 
To tackle this issue, we adopt an iterative method \cite{chu2020robust}.
By taking $\mathrm{tr}\left ( \mathbf{W}_{c} \right ) - \lambda _{\max}\left ( \mathbf{W}_{c} \right )$ as an example,
we have following inequality
\begin{align}
    \mathrm{tr}\small ( \mathbf{W}_{c} \small ) - \small (\mathbf{v}_{c,\max}^{\left [ n \right ]}   \small )^{H} \mathbf{W}_{c} \mathbf{v}_{c,\max}^{\left [ n \right ]} \geq
\mathrm{tr}\small ( \mathbf{W}_{c} \small ) - \lambda _{\max}\small ( \mathbf{W}_{c} \small ) \geq 0,
\label{lower-bound}
\end{align}
where $\mathbf{v}_{c,\max}$ is the normalized eigenvector corresponding to the maximum eigenvalue $\lambda _{\max}\left ( \mathbf{W}_{c} \right )$.
Furthermore, we define $\mathbf{v}_{n_{s},\max}$ as the corresponding eigenvector of $\lambda _{\max}\left ( \mathbf{W}_{n_{s}} \right )$, and so does $\mathbf{b}_{c,\max}$ for $\lambda _{\max}\left ( \mathbf{P}_{c} \right )$ and $\mathbf{b}_{k_{t},\max}$ for $\lambda _{\max}\left ( \mathbf{P}_{k_{t}} \right )$.
Let $\mathrm{PF} $ denote the iterative penalty function
\begin{align}
\mathrm{PF} 
& =\beta \Big (\big [\mathrm{tr}\left ( \mathbf{W}_{c} \right ) - \left (\mathbf{v}_{c,\max}^{\left [ n \right ]}   \right )^{H} \mathbf{W}_{c} \mathbf{v}_{c,\max}^{\left [ n \right ]}  \big ] 
+ \sum_{n_{s}=1}^{N_{s}}\big [\mathrm{tr}\left ( \mathbf{W}_{n_{s}} \right ) - \left (\mathbf{v}_{n_{s},\max}^{\left [ n \right ]}   \right )^{H} \mathbf{W}_{n_{s}} \mathbf{v}_{n_{s},\max}^{\left [ n \right ]}  \big ] \notag
\\ 
& + \big [\mathrm{tr}\left ( \mathbf{P}_{c} \right ) - \left (\mathbf{b}_{c,\max}^{\left [ n \right ]}   \right )^{H} \mathbf{P}_{c} \mathbf{b}_{c,\max}^{\left [ n \right ]}  \big ] 
+ \sum_{k_{t}=1}^{K_{t}}\big [\mathrm{tr}\left ( \mathbf{P}_{k_{t}} \right ) - \left (\mathbf{b}_{k_{t},\max}^{\left [ n \right ]}   \right )^{H} \mathbf{P}_{c} \mathbf{b}_{k_{t},\max}^{\left [ n \right ]} \big ]   \Big ).
\label{q118}
\end{align}
Then, the problem $\mathcal{F}_{1}$ can be reformulated as
\begin{align}
\mathcal{G}_{1}: \  & \max_{W,P,\mathbf{c}^{sat},\mathbf{c}^{bs},q,\mathbf{r},\mathbf{\alpha}, \eta, \xi, t}  q  - \mathrm{PF} 
\label{g1_objective}
\\
s.t. \ 
& (\ref{p1_constraint2}),
(\ref{p1_constraint5}) ,
(\ref{e1_constraint1}),
(\ref{e1_constraint3}) ,
(\ref{d1_constraint1})
-(\ref{d1_constraint4}),
(\ref{f1_constraint1}),
(\ref{f1_constraint4}),
(\ref{f1_constraint7}),
(\ref{f1_constraint10})
\notag
\\
& (\ref{g1_constraint1}),(\ref{g1_constraint3}),(\ref{g1_constraint5}),(\ref{g1_constraint6}),(\ref{g1_constraint9}),(\ref{g1_constraint10}),\ \forall k_{t}\in \mathcal{K}_{t} \notag
\\
& (\ref{g1_constraint2}),(\ref{g1_constraint4}),(\ref{g1_constraint7}),(\ref{g1_constraint8}),(\ref{g1_constraint11}),(\ref{g1_constraint12}),\ \forall k_{s}\in \mathcal{K}_{s} \notag
\end{align}

Now, the problem is convex and can be effectively solved by the CVX toolbox.
The results
$\small (  W^{\left [ n \right ]},P^{\left [ n \right ]},\eta^{\left [ n \right ]},\xi^{\left [ n \right ]},t^{\left [ n \right ]} \small)$ obtained from the $n$-th iteration are treated as constants while solving $\mathcal{G}_{1}$.
The objective function is guaranteed to converge by the existence of lower bounds, i.e., (\ref{lower-bound}).
In other words, the rank-one constraints can be satisfied \cite{chu2020robust}.
We summarize the procedure of this robust joint beamforming scheme in Algorithm \ref{algorithm2}.
Finally, eigenvalue decomposition (EVD) can be used to obtain the optimized beamforming vectors of both satellite and BS.
The MMF optimization problem for RSMA-based \textit{cooperative STIN} with satellite phase uncertainty taken into account remains the same as $\mathcal{P}_{2}$ in Section III.
Here,  we  still omit  the detailed optimization framework.
The process keeps the same as that for the \textit{coordinated STIN} scenario.

\begin{algorithm}
\caption{Robust Joint Beamforming Scheme}\label{algorithm2}
\begin{algorithmic}
\State \textbf{Initialize}: $n\leftarrow 0,W^{\left [ n \right ]},P^{\left [ n \right ]},\eta^{\left [ n \right ]},\xi^{\left [ n \right ]},t^{\left [ n \right ]}$;
\Repeat
\State Solve $\mathcal{G}_{1}$ at 
$\small (  W^{\left [ n \right ]},P^{\left [ n \right ]},\eta^{\left [ n \right ]},\xi^{\left [ n \right ]},t^{\left [ n \right ]} \small)$ to get

the optimal solution
$\small (  \breve{W},\breve{P},\breve{\eta},\breve{\xi},\breve{t},\breve{\mathrm{objective}} \small )$;
\State $n\leftarrow n+1$; 
\State Update $ W^{\left [ n\right ]}\leftarrow \breve{W},P^{\left [ n \right ]}\leftarrow \breve{P},\eta^{\left [ n \right ]}\leftarrow \breve{\eta},\xi^{\left [ n \right ]}\leftarrow \breve{\xi},t^{\left [ n \right ]}\leftarrow \breve{t}, 
\mathrm{objective}^{\left [ n\right ]}\leftarrow \breve{\mathrm{objective}};$
\Until{$\big |\mathrm{objective}^{\left [ n \right ]}-\mathrm{objective}^{\left [ n -1\right ]}  \big |< \varepsilon$ ;}
\end{algorithmic}
\end{algorithm}

\section{Simulation Results}

In this section, 
simulation results are provided to evaluate the performance
of our proposed joint beamforming algorithms.
Both perfect CSIT and imperfect CSIT with satellite channel phase uncertainties are considered.
The tolerance of accuracy is set to be $\varepsilon = 10^{-5}$.
Channel models have been introduced in Section II, and the simulation parameters are listed in Table \ref{table_example}.
Specifically, we assume the satellite is equipped with $N_{s} = 3$ antennas, and $\rho = 2$ multicasting SUs locate uniformly in each beam coverage area. 
According to the architecture of SFPB, the number of SUs is $K_{s}= \rho N_{s} =6$. 
Meanwhile, the BS is deployed with $N_{t} = 4 \times 4 = 16$ UPAs, and there are $K_{t} = 4$ CUs uniformly distributed within the BS coverage. 
Since the noise power is normalized by $\kappa T_{sys}B_{w} $ in (\ref{noise_norm}), we set unit noise variance, i.e., $\sigma _{k_{s}}^{sat2} = \sigma _{k_{t}}^{bs2}=1,\ \forall k_{s} \in \mathcal{K}_{s},\ \forall k_{t} \in \mathcal{K}_{t}$.
All MMF rate curves throughout the simulation are calculated by averaging $100$ channel realizations.

\begin{figure}
\centering
\begin{minipage}[t]{0.49\textwidth}
\centering
\includegraphics[width=8.5cm]{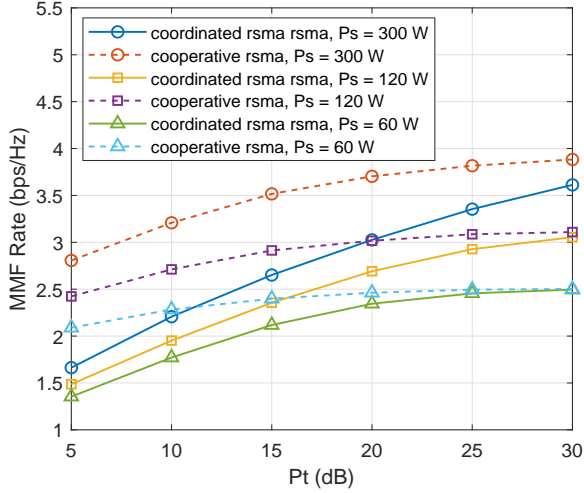}
\caption{MMF rate versus $P_{t}$ for different $P_{s}$ setups. 
$N_{t}=16$, $K_{t} = 4$, $N_{s}=3$, $K_{s} = 6$.}
\label{fig:figure2}
\end{minipage}
\begin{minipage}[t]{0.49\textwidth}
\centering
\includegraphics[width=8.5cm]{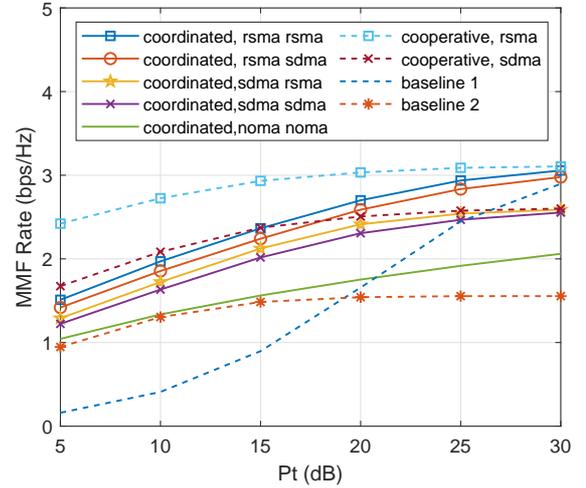}
\caption{MMF rate versus $P_{t}$ for different transmission strategies. 
$N_{t}=16$, $K_{t} = 4$, $N_{s}=3$, $K_{s} = 6$, $P_{s} = 120 \mathrm{W}$. }
\label{fig:figure3}
\end{minipage}
\end{figure}

At first, we assume that perfect CSI is available at the GW.
Fig. \ref{fig:figure2}
compare the MMF rate performance of RSMA-based \textit{coordinated} and \textit{cooperative scheme}.
The label ``coordinated rsma rsma" means RSMA is adopted at both the satellite and BS,
while ``cooperative rsma" means the satellite and BS work cooperatively as a super transmitter and RSMA is adopted.
As the BS transmit power budget $P_{t}$ grows, we can see that the MMF rates of both schemes increase and tend to saturate at large $P_{t}$ region.
The \textit{cooperative scheme} outperforms the \textit{coordinated scheme} apparently at low $P_{t}$ region. 
The gap between two schemes decreases gradually as $P_{t}$ grows, and finally converge to the same value when $P_{t}$ is sufficiently large.
The reasons are as follows. 
When $P_{t}$ is relatively small, the STIN’s performance is restricted in the \textit{coordinated scheme} because the SINRs of CUs are much lower than the SINRs of SUs. 
Joint beamforming is designed to achieve optimal MMF rates.
However, in the \textit{cooperative scheme}, transmit data sharing is assumed  and the satellite can complement the services of BS to CUs, thereby remaining the optimized MMF rate at a higher level than that in the \textit{coordinated scheme}.
As $P_{t}$ grows, the benefit of 
the \textit{cooperative scheme} compared with \textit{coordinated scheme} decreases. Finally, when $P_{t}$ is sufficiently large, the MMF rates of both schemes are restricted and converge to the save value due to the fixed satellite transmit power budget $P_{s}$. 
We also investigate the influence of different $P_{s}$ values.
Apparently, the larger $P_{s}$ is, the better MMF rate performance can be achieved.

\begin{table} 
\vspace{-1.5em}
\caption{Simulation Parameters}
\label{table_example}
\centering
\begin{tabular}{c|c}
\hline
\textbf{Parameter} & \textbf{Value}\\
\hline
Frequency band (carrier frequency) & Ka $\left (28\ \mathrm{GHz}  \right )$\\
Satellite height & $35786\ \mathrm{km}\left ( \mathrm{GEO }\right )$\\
Bandwidth & 500 MHz\\
3 dB angle & $0.4\degree$ \\
Maximum beam gain & 52 dBi\\
User terminal antenna gain & 42.7 dBi\\
System noise temperature &300 K\\
Rain fading parameters & $\left ( \mu,\sigma  \right )=\left ( -3.125,1.591 \right )$\\
UPA inter-element spacing & $d_{1}=d_{2} = \frac{\lambda}{2}$\\
Number of NLoS paths & $3$\\
\hline
\end{tabular}
\end{table}

Fig. \ref{fig:figure3} depicts the MMF rates versus $P_{t}$ for different transmission strategies.
Here, we fix $P_{s}$
 to  be $120 \ \mathrm{Watts}$.
Both RSMA and conventional SDMA are considered in the \textit{cooperative scheme}.
SDMA is a special case of RSMA which can be simulated by discarding the common stream and allocating all transmit power to the private streams.
Different combinations of transmission strategies are considered in the \textit{coordinated scheme}, e.g., the label ``coordinated rsma sdma" means RSMA is exploited at the satellite while SDMA is used at the BS.
It has been shown in \cite{yin2020rate} that adopting RSMA compared with SDMA in an overloaded system can provide more gains than in an underloaded system.
Therefore, in this STIN, the performance improvement obtained by using RSMA compared with SDMA at the satellite is more obvious than at the BS.
As a consequence, the ``coordinated rsma rsma" successively outperforms
``coordinated rsma sdma", ``coordinated sdma rsma" and ``coordinated sdma sdma" .
For the ``coordinated noma noma", SC-SIC is implemented at both the satellite and BS. 
The decoding order of NOMA at the satellite is decided by the ascending order of the weakest user's channel strength in each beam.
We can observe that the MMF rate achieved by NOMA is the worst compared with RSMA and SDMA.
The low performance of NOMA in multi-antenna settings is inline
with the observations in \cite{clerckx2021noma} and the references therein.
As discussed in Fig. \ref{fig:figure2}, \textit{cooperative schemes} can provide higher MMF rates than the corresponding \textit{coordinated schemes}.
Thus, the ``cooperative rsma" outperforms  ``coordinated rsma rsma" in Fig. \ref{fig:figure3}, and finally they tend to reach the same MMF rate restricted by the fixed $P_{s}$ at very large $P_{t}$ region.
Similarly,
the ``cooperative sdma" outperforms  ``coordinated sdma sdma".
As $P_{t}$ increases, they converge to the same value which is lower than that of RSMA.
We do not consider ``cooperative noma" in this STIN because NOMA leads to not only a waste of spatial resources (and therefore rate loss) in multi-antenna settings but also additional decoding order selection complexity and large receiver complexity \cite{clerckx2021noma}.

Two baseline scenarios are also evaluated.
``Baseline 1" represents a two-step beamforming design.
Neither CSI or data information is exchanged between the satellite and BS.
The beamforming for the satellite is at first optimized. 
Then, the beamforming for the BS is optimized.
We consider RSMA at both the satellite and the BS.
Based on such two-step beamfoming, we calculate the minimum rate among all SUs and CUs. 
The rate curve is still generated by averaging $100$ channel realizations. 
Since the satellite beamfoming vectors are not jointly designed with the BS beamforming vectors, CUs will see serious interference from the satellite.
At low $P_{t}$ region, the extremely poor minimum rate performance comes from the rates of CUs.
At high $P_{t}$ region, the rates of CUs are relatively large even if they see severe interference from the satellite, and the minimum rate is decided by the rates of SUs.
Therefore, as $P_{t}$ grows, the value of minimum rate tends to reach the saturation MMF rate of RSMA-based \textit{coordinated} and \textit{cooperative schemes}.
In ``Baseline 2", the satellite and BS operate on different frequency bands.
We still consider RSMA at both the satellite and BS.
This traditional baseline strategy cannot effectively make use of the spectrum, and thus results in poor MMF rate performance.

\begin{figure}
\centering
\begin{minipage}[t]{0.49\textwidth}
\centering
\includegraphics[width=8.5cm]{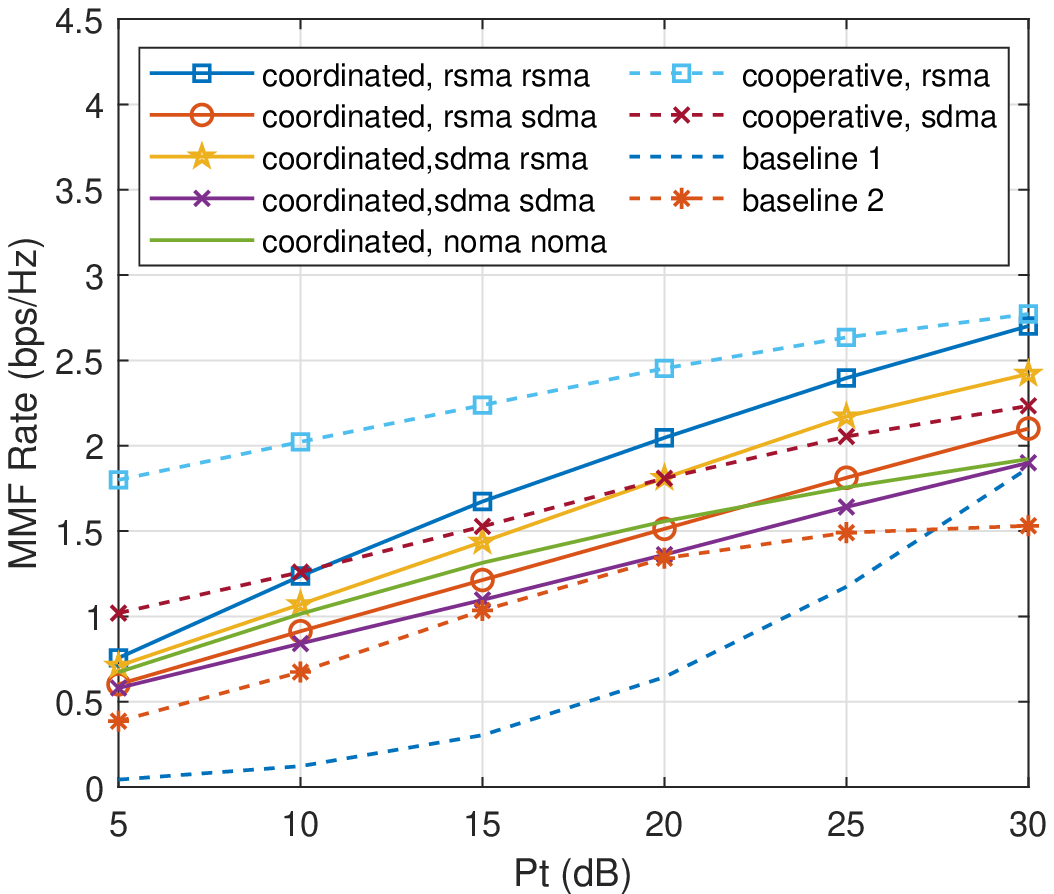}
\caption{MMF rate versus $P_{t}$ for different transmission strategies.
$N_{t}=4$, $K_{t} = 4$, $N_{s}=3$, $K_{s} = 6$, $P_{s} = 120 \mathrm{W}$. }
\label{fig:figure4}
\end{minipage}
\begin{minipage}[t]{0.49\textwidth}
\centering
\includegraphics[width=8.5cm]{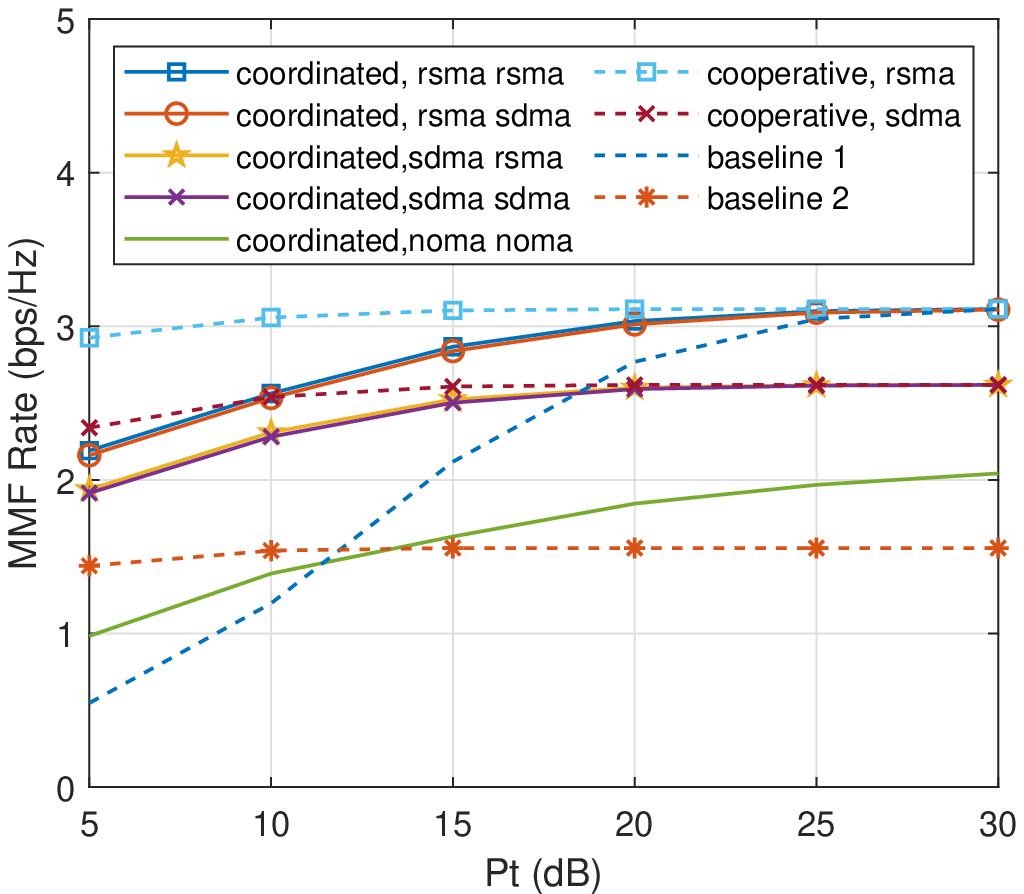}
\caption{MMF rate versus $P_{t}$ for different transmission strategies. 
$N_{t}=64$, $K_{t} = 4$, $N_{s}=3$, $K_{s} = 6$, $P_{s} = 120 \mathrm{W}$. }
\label{fig:figure5}
\end{minipage}
\end{figure}

Fig. \ref{fig:figure4} and Fig. \ref{fig:figure5} illustrate the influence of the number of BS transmit antennas.
In Fig. \ref{fig:figure4}, we set $N_{t} = 2 \times 2 = 4$ UPAs, which is not enough to support effective beamforming at the BS so as to mitigate the intra-cell interference and the interference from the satellite.
Compared with Fig. \ref{fig:figure3} with $16$ UPAs, the MMF rates of all strategies are suppressed.
Specifically, the performance of ``coordinated sdma rsma" becomes better than the ``coordinated rsma sdma". 
It implies that when $N_{t}$ is not sufficient to suppress the interference suffered by the CUs, the gains obtained by using RSMA compared with SDMA at the BS can become more obvious than at the satellite.
In Fig. \ref{fig:figure5}, we assume $N_{t} = 8 \times 8 = 64$ UPAs, which is sufficiently large to support effective beamforming at the BS to eliminate the interference suffered by the CUs.
The benefit of using RSMA compared with SDMA at the BS can be ignored as the terrestrial sub-network is adequately underloaded.
As a result, the curve of ``coordinated rsma rsma" and ``coordinated rsma sdma" overlap, and so do the the curve of ``coordinated sdma rsma" and ``coordinated sdma sdma".
By comparing Fig. \ref{fig:figure3}, Fig. \ref{fig:figure4} and Fig. \ref{fig:figure5}, we can conclude that the larger $N_{t}$ is, the better MMF rate performance can be achieved,
however different setups of $N_{t}$ will not affect the saturation values of each transmission strategy.
In other words, as $N_{t}$ increases, less $P_{t}$ is required to reach the same MMF rate performance. 

\begin{figure}
\centering
\begin{minipage}[t]{0.48\textwidth}
\centering
\includegraphics[width=8.5cm]{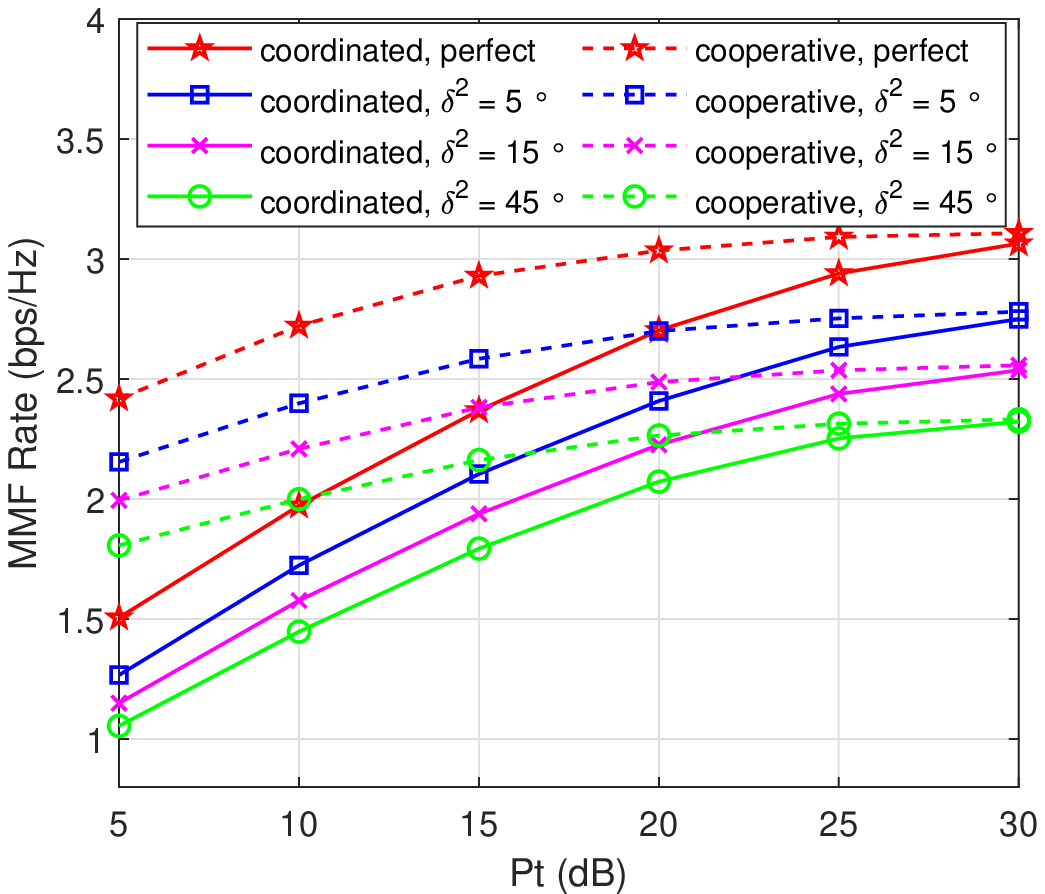}
\caption{MMF rate versus $P_{t}$ for different satellite phase uncertainties. RSMA is used at the transmitters.
$N_{t}=16$, $K_{t} = 4$, $N_{s}=3$, $K_{s} = 6$, $P_{s} = 120 \mathrm{W}$. }
\label{fig:figure6}
\end{minipage}
\begin{minipage}[t]{0.48\textwidth}
\centering
\includegraphics[width=8.5cm]{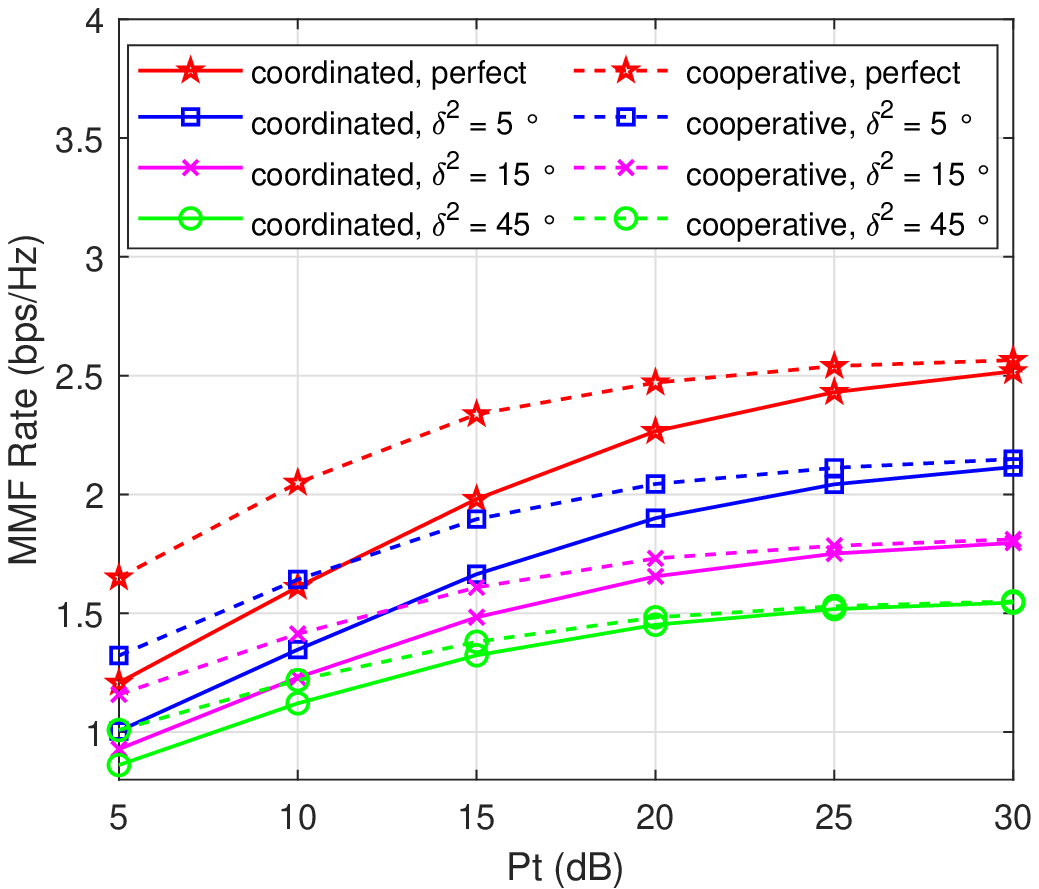}
\caption{MMF rate versus $P_{t}$ for different satellite phase uncertainties.
SDMA is used at the transmitters.
$N_{t}=16$, $K_{t} = 4$, $N_{s}=3$, $K_{s} = 6$, $P_{s} = 120 \mathrm{W}$. }
\label{fig:figure7}
\end{minipage}
\end{figure}

Furthermore, we assume imperfect satellite CSI at the GW with phase uncertainties.
Fig. \ref{fig:figure6} shows the MMF rate performance of the proposed robust joint beamforming scheme in both RSMA-based \textit{coordinated STIN} and \textit{cooperative STIN}.
As the variance of phase uncertainty $\delta^{2}$ increases, the MMF rates of both scenarios decrease gradually. 
When $P_{t} = 20\  \mathrm{dB}$,
from perfect CSIT to the assumptions when $\delta^{2}=5 ^{\circ}$, $\delta^{2}=15 ^{\circ}$, and $\delta^{2}=45 ^{\circ}$, the corresponding MMF rate decreases by $10.74 \%$, $17.50 \%$, $23.22 \%$ respectively in the \textit{coordinated STIN},
and decreases by
$10.37 \%$, $17.69 \%$, $25.04 \%$ respectively in the \textit{cooperative STIN}.
For comparison, SDMA is adopted at the transmitters, i.e.,
we consider the relatively competitive transmission strategy ``cooperative sdma" and ``coordinated sdma sdma" to evaluate the MMF rate performance with satellite channel phase uncertainty.
From Fig.\ref{fig:figure7},
we can observe that the gaps between perfect CSIT curves and imperfect CSIT curves become larger compared with the RSMA results in Fig.\ref{fig:figure6}.
When $P_{t} = 20\  \mathrm{dB}$,
from perfect CSIT to the assumptions when $\delta^{2}=5 ^{\circ}$, $\delta^{2}=15 ^{\circ}$, and $\delta^{2}=45 ^{\circ}$, the corresponding MMF rate decreases by $16.14 \%$, $26.96 \%$, $35.94 \%$ respectively in the \textit{coordinated STIN},
and decreases by
$16.29 \%$, $29.41 \%$, $39.18 \%$ respectively in the \textit{cooperative STIN}.
Therefore, RSMA is more robust to the channel phase uncertainty than SDMA due to its more flexible architecture to partially decoding the interference and partially treat the interference as noise.

\section{Conclusion}
In this work, we investigate the application of RSMA to STIN considering either perfect CSI or imperfect CSI with satellite channel phase uncertainties at the GW.
Two RSMA-based STIN schemes are presented, namely the \textit{coordinated scheme} relying on CSI sharing and the \textit{cooperative scheme} relying on both CSI and data sharing.
MMF optimization problems are 
formulated while satisfying transmit power budgets at both the BS and the satellite.
To tackle the optimization,
two iterative algorithms are respectively proposed for the joint beamforming design considering either perfect or imperfect CSIT.
Through simulation results,
the superiority of  our  proposed  RSMA-based schemes  for STIN is demonstrated compared with various benchmark strategies.
The robustness of RSMA is verified in the presence of channel  phase  uncertainties.
In conclusion, RSMA is shown very promising for STINs to manage the interference in and between the satellite and terrestrial sub-networks.

\ifCLASSOPTIONcaptionsoff
  \newpage
\fi

\bibliographystyle{IEEEtran}
\bibliography{reference}

\begin{thebibliography}{10}
\providecommand{\url}[1]{#1}
\csname url@samestyle\endcsname
\providecommand{\newblock}{\relax}
\providecommand{\bibinfo}[2]{#2}
\providecommand{\BIBentrySTDinterwordspacing}{\spaceskip=0pt\relax}
\providecommand{\BIBentryALTinterwordstretchfactor}{4}
\providecommand{\BIBentryALTinterwordspacing}{\spaceskip=\fontdimen2\font plus
\BIBentryALTinterwordstretchfactor\fontdimen3\font minus
  \fontdimen4\font\relax}
\providecommand{\BIBforeignlanguage}[2]{{%
\expandafter\ifx\csname l@#1\endcsname\relax
\typeout{** WARNING: IEEEtran.bst: No hyphenation pattern has been}%
\typeout{** loaded for the language `#1'. Using the pattern for}%
\typeout{** the default language instead.}%
\else
\language=\csname l@#1\endcsname
\fi
#2}}
\providecommand{\BIBdecl}{\relax}
\BIBdecl

\bibitem{kawamoto2014prospects}
Y.~Kawamoto, Z.~M. Fadlullah, H.~Nishiyama, N.~Kato, and M.~Toyoshima,
  ``Prospects and challenges of context-aware multimedia content delivery in
  cooperative satellite and terrestrial networks,'' \emph{IEEE Communications
  Magazine}, vol.~52, no.~6, pp. 55--61, 2014.

\bibitem{zhang2019multicast}
H.~Zhang, C.~Jiang, J.~Wang, L.~Wang, Y.~Ren, and L.~Hanzo, ``Multicast
  beamforming optimization in cloud-based heterogeneous terrestrial and
  satellite networks,'' \emph{IEEE Transactions on Vehicular Technology},
  vol.~69, no.~2, pp. 1766--1776, 2019.

\bibitem{maleki2015cognitive}
S.~Maleki, S.~Chatzinotas, B.~Evans, K.~Liolis, J.~Grotz, A.~Vanelli-Coralli,
  and N.~Chuberre, ``Cognitive spectrum utilization in ka band multibeam
  satellite communications,'' \emph{IEEE Communications Magazine}, vol.~53,
  no.~3, pp. 24--29, 2015.

\bibitem{lin2018joint}
M.~Lin, Z.~Lin, W.-P. Zhu, and J.-B. Wang, ``Joint beamforming for secure
  communication in cognitive satellite terrestrial networks,'' \emph{IEEE
  Journal on Selected Areas in Communications}, vol.~36, no.~5, pp. 1017--1029,
  2018.

\bibitem{vassaki2013power}
S.~Vassaki, M.~I. Poulakis, A.~D. Panagopoulos, and P.~Constantinou, ``Power
  allocation in cognitive satellite terrestrial networks with qos
  constraints,'' \emph{IEEE Communications Letters}, vol.~17, no.~7, pp.
  1344--1347, 2013.

\bibitem{jia2016broadband}
M.~Jia, X.~Gu, Q.~Guo, W.~Xiang, and N.~Zhang, ``Broadband hybrid
  satellite-terrestrial communication systems based on cognitive radio toward
  {5G},'' \emph{IEEE Wireless Communications}, vol.~23, no.~6, pp. 96--106,
  2016.

\bibitem{choi2015challenges}
J.~P. Choi and C.~Joo, ``Challenges for efficient and seamless
  space-terrestrial heterogeneous networks,'' \emph{IEEE Communications
  Magazine}, vol.~53, no.~5, pp. 156--162, 2015.

\bibitem{sharma2013transmit}
S.~K. Sharma, S.~Chatzinotas, and B.~Ottersten, ``Transmit beamforming for
  spectral coexistence of satellite and terrestrial networks,'' in \emph{8th
  International Conference on Cognitive Radio Oriented Wireless
  Networks}.\hskip 1em plus 0.5em minus 0.4em\relax IEEE, 2013, pp. 275--281.

\bibitem{an2016outage}
K.~An, M.~Lin, W.-P. Zhu, Y.~Huang, and G.~Zheng, ``Outage performance of
  cognitive hybrid satellite--terrestrial networks with interference
  constraint,'' \emph{IEEE Transactions on Vehicular Technology}, vol.~65,
  no.~11, pp. 9397--9404, 2016.

\bibitem{zhu2018cooperative}
X.~Zhu, C.~Jiang, L.~Yin, L.~Kuang, N.~Ge, and J.~Lu, ``Cooperative multigroup
  multicast transmission in integrated terrestrial-satellite networks,''
  \emph{IEEE Journal on Selected Areas in Communications}, vol.~36, no.~5, pp.
  981--992, 2018.

\bibitem{zhang2019joint}
Y.~Zhang, L.~Yin, C.~Jiang, and Y.~Qian, ``Joint beamforming design and
  resource allocation for terrestrial-satellite cooperation system,''
  \emph{IEEE Transactions on Communications}, vol.~68, no.~2, pp. 778--791,
  2019.

\bibitem{lin2018joint1}
Z.~Lin, M.~Lin, J.-B. Wang, X.~Wu, and W.-P. Zhu, ``Joint optimization for
  secure wipt in satellite-terrestrial integrated networks,'' in \emph{2018
  IEEE Global Communications Conference (GLOBECOM)}.\hskip 1em plus 0.5em minus
  0.4em\relax IEEE, 2018, pp. 1--6.

\bibitem{li2020energy}
J.~Li, K.~Xue, D.~S. Wei, J.~Liu, and Y.~Zhang, ``Energy efficiency and traffic
  offloading optimization in integrated satellite/terrestrial radio access
  networks,'' \emph{IEEE Transactions on Wireless Communications}, vol.~19,
  no.~4, pp. 2367--2381, 2020.

\bibitem{sharma2020secure}
P.~K. Sharma and D.~I. Kim, ``Secure 3d mobile uav relaying for hybrid
  satellite-terrestrial networks,'' \emph{IEEE Transactions on Wireless
  Communications}, vol.~19, no.~4, pp. 2770--2784, 2020.

\bibitem{lu2019robust}
W.~Lu, K.~An, and T.~Liang, ``Robust beamforming design for sum secrecy rate
  maximization in multibeam satellite systems,'' \emph{IEEE Transactions on
  Aerospace and Electronic Systems}, vol.~55, no.~3, pp. 1568--1572, 2019.

\bibitem{guo2019robust}
X.~Guo, D.~Yang, Z.~Luo, H.~Wang, and J.~Kuang, ``Robust {THP} design for
  energy efficiency of multibeam satellite systems with imperfect {CSI},''
  \emph{IEEE Communications Letters}, vol.~24, no.~2, pp. 428--432, 2019.

\bibitem{lin2019robust}
Z.~Lin, M.~Lin, J.~Ouyang, W.-P. Zhu, A.~D. Panagopoulos, and M.-S. Alouini,
  ``Robust secure beamforming for multibeam satellite communication systems,''
  \emph{IEEE Transactions on Vehicular Technology}, vol.~68, no.~6, pp.
  6202--6206, 2019.

\bibitem{gharanjik2015robust}
A.~Gharanjik, M.~B. Shankar, P.-D. Arapoglou, M.~Bengtsson, and B.~Ottersten,
  ``Robust precoding design for multibeam downlink satellite channel with phase
  uncertainty,'' in \emph{2015 IEEE International Conference on Acoustics,
  Speech and Signal Processing (ICASSP)}.\hskip 1em plus 0.5em minus
  0.4em\relax IEEE, 2015, pp. 3083--3087.

\bibitem{wang2021resource}
W.~Wang, L.~Gao, R.~Ding, J.~Lei, L.~You, C.~A. Chan, and X.~Gao, ``Resource
  efficiency optimization for robust beamforming in multi-beam satellite
  communications,'' \emph{IEEE Transactions on Vehicular Technology}, 2021.

\bibitem{chu2020robust}
J.~Chu, X.~Chen, C.~Zhong, and Z.~Zhang, ``Robust design for noma-based
  multibeam {LEO} satellite internet of things,'' \emph{IEEE Internet of Things
  Journal}, vol.~8, no.~3, pp. 1959--1970, 2020.

\bibitem{vazquez2016precoding}
M.~{\'A}. V{\'a}zquez, A.~Perez-Neira, D.~Christopoulos, S.~Chatzinotas,
  B.~Ottersten, P.-D. Arapoglou, A.~Ginesi, and G.~Taricco, ``Precoding in
  multibeam satellite communications: Present and future challenges,''
  \emph{IEEE Wireless Communications}, vol.~23, no.~6, pp. 88--95, 2016.

\bibitem{DVB-S2}
``{Second generation framing structure, channel coding modulation systems
  broadcasting, interactive services, news gathering other broad-band satellite
  applications (DVB-S2)},'' European Broadcasting Union (EBU), document ETSI EN
  302 307–1 V1.4.1, Standard, Nov. 2014.

\bibitem{DVB-S2x}
``{Second generation framing structure, channel coding and modulation systems
  for broadcasting, interative services, news gathering and other broadband
  satellite applications; Part 2: DVB-S2 Extensions (DVB-S2X)},'' European
  Broadcasting Union (EBU), document ETSI EN 302-307-2 V1.1.1, Standard, Oct.
  2014.

\bibitem{clerckx2019rate}
B.~Clerckx, Y.~Mao, R.~Schober, and H.~V. Poor, ``Rate-splitting unifying
  {SDMA}, {OMA}, {NOMA}, and multicasting in {MISO} broadcast channel: A simple
  two-user rate analysis,'' \emph{IEEE Wireless Communications Letters},
  vol.~9, no.~3, pp. 349--353, 2019.

\bibitem{mao2018rate}
Y.~Mao, B.~Clerckx, and V.~O. Li, ``Rate-splitting multiple access for downlink
  communication systems: bridging, generalizing, and outperforming {SDMA} and
  {NOMA},'' \emph{EURASIP journal on wireless communications and networking},
  vol. 2018, no.~1, p. 133, 2018.

\bibitem{ahmad2019interference}
A.~A. Ahmad, H.~Dahrouj, A.~Chaaban, A.~Sezgin, and M.-S. Alouini,
  ``Interference mitigation via rate-splitting and common message decoding in
  cloud radio access networks,'' \emph{IEEE Access}, vol.~7, pp.
  80\,350--80\,365, 2019.

\bibitem{zhang2019cooperative}
J.~Zhang, B.~Clerckx, J.~Ge, and Y.~Mao, ``Cooperative rate-splitting for
  {MISO} broadcast channel with user relaying, and performance benefits over
  cooperative {NOMA},'' \emph{IEEE Signal Processing Letters}, vol.~26, no.~11,
  pp. 1678--1682, 2019.

\bibitem{joudeh2016sum}
H.~Joudeh and B.~Clerckx, ``Sum-rate maximization for linearly precoded
  downlink multiuser {MISO} systems with partial {CSIT}: A rate-splitting
  approach,'' \emph{IEEE Transactions on Communications}, vol.~64, no.~11, pp.
  4847--4861, 2016.

\bibitem{hao2015rate}
C.~Hao, Y.~Wu, and B.~Clerckx, ``Rate analysis of two-receiver {MISO} broadcast
  channel with finite rate feedback: A rate-splitting approach,'' \emph{IEEE
  Transactions on Communications}, vol.~63, no.~9, pp. 3232--3246, 2015.

\bibitem{joudeh2016robust}
H.~Joudeh and B.~Clerckx, ``Robust transmission in downlink multiuser {MISO}
  systems: A rate-splitting approach,'' \emph{IEEE Transactions on Signal
  Processing}, vol.~64, no.~23, pp. 6227--6242, 2016.

\bibitem{piovano2017optimal}
E.~Piovano and B.~Clerckx, ``Optimal {DoF} region of the $ k $-user {MISO BC}
  with partial {CSIT},'' \emph{IEEE Communications Letters}, vol.~21, no.~11,
  pp. 2368--2371, 2017.

\bibitem{mao2020beyond}
Y.~Mao and B.~Clerckx, ``Beyond dirty paper coding for multi-antenna broadcast
  channel with partial {CSIT}: A rate-splitting approach,'' \emph{IEEE
  Transactions on Communications}, vol.~68, no.~11, pp. 6775--6791, 2020.

\bibitem{joudeh2017rate}
H.~Joudeh and B.~Clerckx, ``Rate-splitting for max-min fair multigroup
  multicast beamforming in overloaded systems,'' \emph{IEEE Transactions on
  Wireless Communications}, vol.~16, no.~11, pp. 7276--7289, 2017.

\bibitem{yalcin2020rate}
A.~Z. Yalcin, M.~Yuksel, and B.~Clerckx, ``Rate splitting for multi-group
  multicasting with a common message,'' \emph{IEEE Transactions on Vehicular
  Technology}, vol.~69, no.~10, pp. 12\,281--12\,285, 2020.

\bibitem{tervo2018multigroup}
O.~Tervo, L.-N. Trant, S.~Chatzinotas, B.~Ottersten, and M.~Juntti,
  ``Multigroup multicast beamforming and antenna selection with rate-splitting
  in multicell systems,'' in \emph{2018 IEEE 19th International Workshop on
  Signal Processing Advances in Wireless Communications (SPAWC)}.\hskip 1em
  plus 0.5em minus 0.4em\relax IEEE, 2018, pp. 1--5.

\bibitem{papazafeiropoulos2017rate}
A.~Papazafeiropoulos, B.~Clerckx, and T.~Ratnarajah, ``Rate-splitting to
  mitigate residual transceiver hardware impairments in massive {MIMO}
  systems,'' \emph{IEEE Transactions on Vehicular Technology}, vol.~66, no.~9,
  pp. 8196--8211, 2017.

\bibitem{dai2017multiuser}
M.~Dai and B.~Clerckx, ``Multiuser millimeter wave beamforming strategies with
  quantized and statistical {CSIT},'' \emph{IEEE Transactions on Wireless
  Communications}, vol.~16, no.~11, pp. 7025--7038, 2017.

\bibitem{mao2019rate}
Y.~Mao, B.~Clerckx, and V.~O. Li, ``Rate-splitting for multi-user multi-antenna
  wireless information and power transfer,'' in \emph{2019 IEEE 20th
  International Workshop on Signal Processing Advances in Wireless
  Communications (SPAWC)}.\hskip 1em plus 0.5em minus 0.4em\relax IEEE, 2019,
  pp. 1--5.

\bibitem{xu2020rate}
C.~Xu, B.~Clerckx, S.~Chen, Y.~Mao, and J.~Zhang, ``Rate-splitting multiple
  access for multi-antenna joint communication and radar transmissions,'' in
  \emph{2020 IEEE International Conference on Communications Workshops (ICC
  Workshops)}.\hskip 1em plus 0.5em minus 0.4em\relax IEEE, 2020, pp. 1--6.

\bibitem{caus2018exploratory}
M.~Caus, A.~Pastore, M.~Navarro, T.~Ram{\'\i}rez, C.~Mosquera, N.~Noels,
  N.~Alagha, and A.~I. Perez-Neira, ``Exploratory analysis of superposition
  coding and rate-splitting for multibeam satellite systems,'' in \emph{2018
  15th International Symposium on Wireless Communication Systems
  (ISWCS)}.\hskip 1em plus 0.5em minus 0.4em\relax IEEE, 2018, pp. 1--5.

\bibitem{vazquez2018rate}
M.~Vazquez, M.~Caus, and A.~Perez-Neira, ``Rate-splitting for {MIMO} multibeam
  satellite systems,'' in \emph{WSA 2018; 22nd International ITG Workshop on
  Smart Antennas}.\hskip 1em plus 0.5em minus 0.4em\relax VDE, 2018, pp. 1--6.

\bibitem{yin2020rate1}
L.~Yin and B.~Clerckx, ``Rate-splitting multiple access for multibeam satellite
  communications,'' in \emph{2020 IEEE International Conference on
  Communications Workshops (ICC Workshops)}.\hskip 1em plus 0.5em minus
  0.4em\relax IEEE, 2020, pp. 1--6.

\bibitem{yin2020rate}
------, ``Rate-splitting multiple access for multigroup multicast and multibeam
  satellite systems,'' \emph{IEEE Transactions on Communications}, vol.~69,
  no.~2, pp. 976--990, 2020.

\bibitem{si2021rate}
Z.~W. Si, L.~Yin, and B.~Clerckx, ``Rate-splitting multiple access for
  multigateway multibeam satellite systems with feeder link interference,''
  \emph{arXiv preprint arXiv:2102.05792}, 2021.

\bibitem{lin2021supporting}
Z.~Lin, M.~Lin, T.~de~Cola, J.-B. Wang, W.-P. Zhu, and J.~Cheng, ``Supporting
  {I}o{T} with rate-splitting multiple access in satellite and aerial
  integrated networks,'' \emph{IEEE Internet of Things Journal}, 2021.

\bibitem{lin2020secure}
Z.~Lin, M.~Lin, B.~Champagne, W.-P. Zhu, and N.~Al-Dhahir, ``Secure and energy
  efficient transmission for {RSMA}-based cognitive satellite-terrestrial
  networks,'' \emph{IEEE Wireless Communications Letters}, vol.~10, no.~2, pp.
  251--255, 2020.

\bibitem{tervo2015optimal}
O.~Tervo, L.-N. Tran, and M.~Juntti, ``Optimal energy-efficient transmit
  beamforming for multi-user {MISO} downlink,'' \emph{IEEE Transactions on
  Signal Processing}, vol.~63, no.~20, pp. 5574--5588, 2015.

\bibitem{nguyen2017distributed}
K.-G. Nguyen, Q.-D. Vu, M.~Juntti, and L.-N. Tran, ``Distributed solutions for
  energy efficiency fairness in multicell {MISO} downlink,'' \emph{IEEE
  Transactions on Wireless Communications}, vol.~16, no.~9, pp. 6232--6247,
  2017.

\bibitem{marks1978general}
B.~R. Marks and G.~P. Wright, ``A general inner approximation algorithm for
  nonconvex mathematical programs,'' \emph{Operations research}, vol.~26,
  no.~4, pp. 681--683, 1978.

\bibitem{gharanjik2015precoding}
A.~Gharanjik, M.~B. Shankar, P.-D. Arapoglou, M.~Bengtsson, and B.~Ottersten,
  ``Precoding design and user selection for multibeam satellite channels,'' in
  \emph{2015 IEEE 16th International Workshop on Signal Processing Advances in
  Wireless Communications (SPAWC)}.\hskip 1em plus 0.5em minus 0.4em\relax
  IEEE, 2015, pp. 420--424.

\bibitem{shao2017simple}
M.~Shao and W.-K. Ma, ``A simple way to approximate average robust multiuser
  miso transmit optimization under covariance-based csit,'' in \emph{2017 IEEE
  International Conference on Acoustics, Speech and Signal Processing
  (ICASSP)}.\hskip 1em plus 0.5em minus 0.4em\relax IEEE, 2017, pp. 3504--3508.

\bibitem{clerckx2021noma}
B.~Clerckx, Y.~Mao, R.~Schober, E.~Jorswieck, D.~J. Love, J.~Yuan, L.~Hanzo,
  G.~Y. Li, E.~G. Larsson, and G.~Caire, ``Is {NOMA} efficient in multi-antenna
  networks? a critical look at next generation multiple access techniques,''
  \emph{IEEE Open Journal of the Communications Society}, 2021.

\end{thebibliography}

\end{document}